\UseRawInputEncoding
\documentclass[11pt]{amsart}
\usepackage[T1]{fontenc}
\usepackage{lmodern}
\usepackage{amsmath,amssymb,color,amsthm,mathtools}
\usepackage{graphicx}
\usepackage[a4paper,margin=28mm]{geometry}
\usepackage{hyperref}
\hypersetup{pdftitle={Dynamics through the first two Lagrange necks in the small-mass restricted three-body problem},pdfauthor={Lei Liu and Pedro A. S. Salomao}}
\newcommand{\R}{\mathbb R}
\newcommand{\C}{\mathbb C}
\newcommand{\Z}{\mathbb Z}
\newcommand{\D}{\mathbb D}
\newcommand{\RP}{\mathbb {RP}}
\newcommand{\h}{\mathfrak h}
\newcommand{\CZ}{\mu_{\rm CZ}}
\newcommand{\A}{\mathcal A}
\newcommand{\slk}{\operatorname{sl}}
\newcommand{\lk}{\operatorname{link}}
\newcommand{\Sp}{\operatorname{Sp}}
\newtheorem{thm}{Theorem}[section]
\newtheorem{prop}[thm]{Proposition}
\newtheorem{lem}[thm]{Lemma}

\theoremstyle{definition}
\newtheorem{defi}[thm]{Definition}

\numberwithin{equation}{section}
\allowdisplaybreaks[2]
\title[The dynamics through two Lagrange necks]{The dynamics through the first two Lagrange necks in the small-mass restricted three-body problem}
\author{Lei Liu}
\address{School of Mathematics, Shandong University}
\email{liulei30@email.sdu.edu.cn}
\author{Pedro A. S. Salom\~ao}
\address{Shenzhen International Center for Mathematics, SUSTech}
\email{psalomao@sustech.edu.cn}
\date{}
\begin{document}
\begin{abstract}
We study the planar circular restricted three-body problem (PCR3BP) for sufficiently small mass ratios, from low energies to slightly above the second Lagrange value. We show that retrograde orbits around the primaries and Lyapunov orbits at the necks form the binding of a finite energy foliation. We construct these foliations using the theory of pseudoholomorphic curves in symplectizations developed by Hofer, Wysocki and Zehnder. Slightly above the second Lagrange value, the dynamics is transverse to families of disks bounded by the retrograde orbits, and cylinders connecting the retrograde orbits to the Lyapunov orbits, forming a $3-2-3-2$ foliation.  The proof uses the rotating Kepler problem and Hill's lunar problem to control the singular limit as the mass ratio $\mu \to 0$. As a consequence of the foliation, both Lyapunov orbits have infinitely many homoclinic orbits.
\end{abstract}
\maketitle
\tableofcontents

\section{Introduction}
The planar circular restricted three-body problem is a classical problem in celestial mechanics. Its dynamics, even for small mass ratios, includes periodic orbits, collisions, transit trajectories, homoclinic and heteroclinic orbits, and chaotic invariant sets. In this paper we consider the following question:
\vspace{0.3cm}

\emph{Is there a simple global structure that organizes the flow of the PCR3BP as the energy passes through the first two Lagrange values?}

\vspace{0.3cm}

For sufficiently small positive mass ratios, we construct such a structure from low energies to slightly above the second Lagrange value. Below $L_1$, the bounded regions around the two primaries are separated. When the energy crosses $L_1$, a neck opens between them. When it crosses $L_2$, a second neck opens on the far side of the Moon and connects the Moon region to the exterior. We construct global foliations that organize the flow in all three cases and fit together with these topological changes.

The periodic orbits that appear in the binding of these foliations are classical ones. Around each primary, there is a retrograde orbit. Each open neck contains a Lyapunov orbit. These periodic orbits form the binding of foliations of the regularized energy levels. The regular leaves are surfaces transverse to the Hamiltonian flow. Below $L_1$, these are global surfaces of section. Between and above the first two Lagrange values, they form finite energy foliations whose leaves are planes and cylinders asymptotic to the binding orbits.

We recall the Hamiltonian formulation of the problem. A massless particle moves under the attraction of two primaries, which move on circular orbits about their center of mass. In rotating coordinates, the Hamiltonian is
\begin{equation}\label{equ: Ham CR3BP}
 H_\mu(p,q)=\frac12\big((p_1-q_2)^2+(p_2+q_1)^2\big)
 -\frac{\mu}{|q-(1-\mu)|}-\frac{1-\mu}{|q+\mu|}-\frac12|q|^2.
\end{equation}
We identify $\R^2$ with $\C$. The primaries at $-\mu$ and $1-\mu$ have masses $1-\mu$ and $\mu$. We call them the Earth and the Moon and assume $0<\mu\leq \frac12$. The Lagrange points are $l_i(\mu)$ and their values are $L_i(\mu)$, ordered by $L_1(\mu)<L_2(\mu)\leq L_3(\mu)<L_4(\mu)=L_5(\mu)$.
The points $l_1$ and $l_2$ lie on opposite sides of the Moon, with $l_1$ facing the Earth.

The local dynamics near a collinear Lagrange point is well known. The equilibrium is of saddle-center type, and the nearby Lyapunov orbit together with its stable and unstable manifolds separates transit from non-transit motion. The local stable and unstable manifolds form cylinders near the neck. Transport through these regions dates back to Conley \cite{Conley68}, and the stable/unstable manifolds of the Lyapunov orbits play a central role in the classical and numerical descriptions of low-energy transfer, see \cite{GomezKoonLoMarsden2004}. This local description does not determine the dynamics on the whole regularized component. Here we study how the necks and the bounded regions around the two primaries are related.

Homoclinic and heteroclinic connections of Lyapunov orbits have been studied by analytical, numerical and computer-assisted methods. Llibre, Mart\'inez and Sim\'o \cite{LlibreMartinezSimo85} analyzed transversality and symbolic dynamics near a collinear libration point in the small-mass regime. Koon, Lo, Marsden and Ross \cite{KoonLoMarsdenRoss00} numerically exhibited heteroclinic connections between Lyapunov orbits around the two necks. Wilczak and Zgliczy\'nski \cite{WilczakZgliczynski03,WilczakZgliczynski05} gave computer assisted proofs of connecting orbits and symbolic dynamics for the Sun--Jupiter--Oterma parameters. Capi\'nski \cite{Capinski12} established validated results on Lyapunov families and transverse invariant-manifold intersections in the Jupiter-Sun problem. Here the connections are obtained from global finite energy foliations throughout the stated small-mass energy ranges.

An additional difficulty comes from the limit of small mass ratios. As $\mu\to 0$, the first two Lagrange values approach each other on the Hill scale. At the same time, the two sides of the problem have different limits. Near the Earth, one finds the rotating Kepler problem, in which transverse foliations have also been studied by Kim \cite{Kim24}. Near the Moon, after the Hill rescaling, one sees Hill's lunar problem \cite{FvK,Hill1}. Neither limit by itself describes the full compact region. We use both limiting systems and obtain uniform index, contact and neck estimates as $L_1$ and $L_2$ approach each other.

We state our main results. Below $L_1$, both bounded regularized components are dynamically convex. In particular, Birkhoff's retrograde orbit bounds a rational disk-like global surface of section. Between $L_1$ and $L_2$, the two bounded components get connected, and we construct a finite energy foliation across the first neck, the so-called $3-2-3$ foliation. Slightly above $L_2$, we construct a finite energy foliation on the compact region bounded by the outer dividing sphere. This is the $3-2-3-2$ foliation, which contains both Lyapunov and retrograde orbits in its binding. These foliations describe the bounded dynamics as the first two necks open.

For energies slightly above the second Lagrange value, the Lyapunov orbits have different actions. The two Lyapunov orbits are no longer dynamically symmetric for small $\mu>0$ and the energy range considered here. We then prove that their actions satisfy $\A(P_{1,2})>\A(P_{2,2})$, where $P_{i,2}$ is near $l_i, i=1,2$.
For the larger-action Lyapunov orbit $P_{1,2}$ near $l_1$, reversibility and the infinite twist near the saddle-center give infinitely many homoclinic orbits and infinitely many periodic orbits. For the Lyapunov orbit $P_{2,2}$ of smaller action, the proof uses the global foliation. Its unstable manifold meets a nearby page in a circle with $\alpha$-integral $\A(P_{2,2})$. We follow this circle through the foliation, allowing transit through the first neck. Its $\alpha$-integral is preserved, so every disk it bounds on the current leaf has $d\alpha$-area $\A(P_{2,2})$. The finite area of the rigid leaves forces a stable intersection, and the action comparison excludes escape through the second neck before such an intersection. Together with reversibility and the infinite twist, this gives infinitely many homoclinic orbits to $P_{2,2}$. In particular, both Lyapunov orbits admit infinitely many homoclinic orbits. Moreover, if the compact region contains a heteroclinic orbit between the two Lyapunov orbits, then it contains infinitely many geometrically distinct transverse heteroclinic orbits in each direction.

\begin{figure}[htbp]
\centering
\makebox[\textwidth][c]{%
\includegraphics[width=1.0\textwidth]{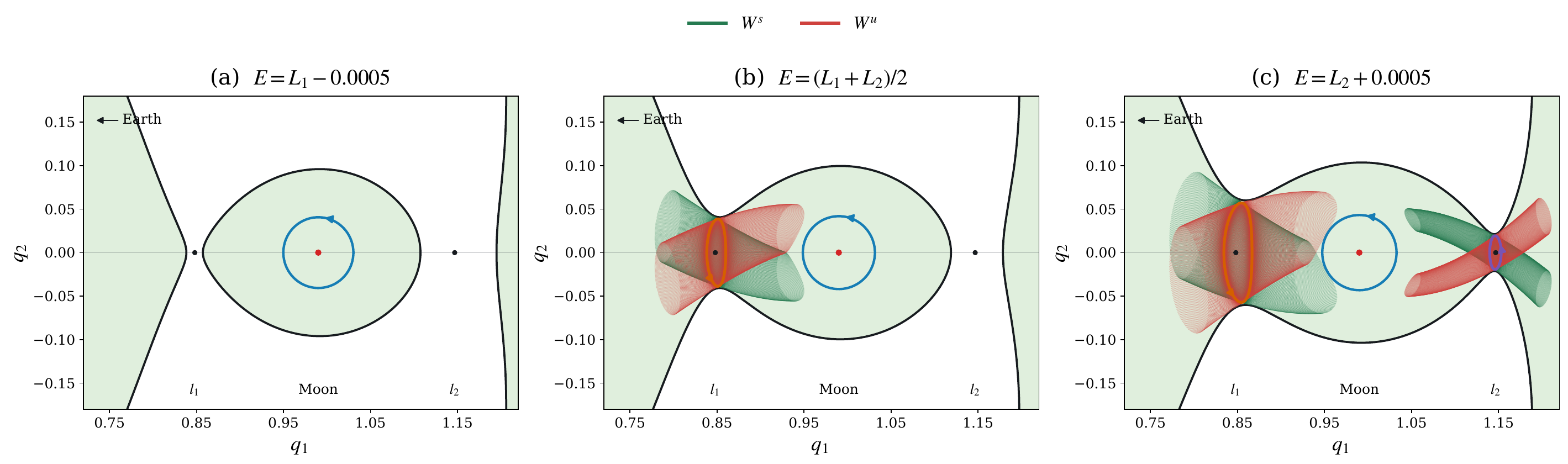}%
}
\caption{Hill regions for $\mu=0.01$ and various energies $E$.
(a) Below $L_1$, the bounded Hill region is disconnected.
(b) Between $L_1$ and $L_2$, the first neck is open.
(c)
Slightly above $L_2$, both necks are open.
The retrograde orbits around the Moon are shown in blue.
The projections of the local stable and unstable manifolds of the
Lyapunov orbits are depicted in green and red, respectively.}
\label{fig:three-energy-regimes}
\end{figure}

The contact geometry of the regularized energy levels of the restricted three-body problem was studied in \cite{AFKP12}, see also \cite{FvK}. Global surfaces of section near the smaller primary in subcritical regimes were obtained by Albers, Fish, Frauenfelder, Hofer and van Koert \cite{AlbersFishFrauenfelderHoferVanKoert12}. A computational approach to Birkhoff's conjecture, including validated calculations of periodic orbits and their Conley--Zehnder indices, is developed in \cite{JoungVanKoert25}. In \cite{LS25} we constructed finite energy foliations for mass ratios close to $\frac12$. In \cite{LS26}, we treated Hill's lunar problem and the first critical level. The present small-mass problem is not a direct perturbation of either paper. The Earth side is governed by the rotating Kepler problem, and the Liouville fields from the two limiting descriptions must be joined near the first neck.

The theory of pseudoholomorphic curves behind the foliations goes back to Hofer, Wysocki, and Zehnder \cite{Hofer93, HWZI, HWZII, HWZIII, HWZ03}. Fast finite energy planes and the moduli-space constructions underlying global surfaces of section were further developed by Hryniewicz \cite{HryFast12, HrySystems14}. Rational global sections and lens-space versions are studied in \cite{HLS15, HS2}. Finite energy foliations for weakly convex Reeb flows were developed in \cite{dPHKS24, dPS1, dPS2, LS25}. For mechanical systems with two degrees of freedom, the authors in \cite{dPKSS26} constructed transverse foliations above critical energies under suitable convexity assumptions, with an application to the H\'enon-Heiles system for energies slightly above $\frac16$. Fish and Siefring studied finite energy foliations under contact connected sum \cite{FishSiefring18}.

\subsection{The first two Lagrange values and the regularized components}

For $E<L_1(\mu)$, the energy surface has two bounded components and one unbounded component. The bounded components project to punctured disks around the primaries. After a collision regularization, each bounded component is diffeomorphic to $\RP^3$. We denote them by $\mathcal M^e_{\mu,E}\equiv \RP^3 \mbox{ and } \mathcal M^m_{\mu,E}\equiv \RP^3$.

For $L_1(\mu)<E<L_2(\mu)$, the two bounded components join through the first neck. The resulting regularized component is $\mathcal M^1_{\mu,E}\cong\RP^3\#\RP^3$.
Its natural double-cover regularization is $S^1\times S^2$. A dividing sphere $\mathcal S_1$ in the neck separates it into two closed chambers, still denoted by $\mathcal M^e_{\mu,E}$ and $\mathcal M^m_{\mu,E}$, each diffeomorphic to $\RP^3$ with an open $3$-ball removed.

For energies above $L_2(\mu)$ and below $L_3(\mu)$, the component containing the primaries is unbounded. We do not compactify its end at infinity. Instead, we cut at a second dividing sphere $\mathcal S_2$ and denote the compact regularized region containing the primaries by $\mathcal M^1_{\mu,E}$. Then
\begin{equation}\label{eq: chamber-topology}
\begin{aligned}
 \mathcal M^1_{\mu,E}&\cong(\RP^3\#\RP^3)\setminus B^3,
 &\partial\mathcal M^1_{\mu,E}&=\mathcal S_2,\\
 \mathcal M^e_{\mu,E}&\cong\RP^3\setminus B^3,
 &\partial\mathcal M^e_{\mu,E}&=\mathcal S_1,\\
 \mathcal M^m_{\mu,E}&\cong\RP^3\setminus(B^3_1\cup B^3_2),
 &\partial\mathcal M^m_{\mu,E}&=\mathcal S_1\cup\mathcal S_2.
\end{aligned}
\end{equation}
The flow crosses the hemispheres of a dividing sphere. Thus, the chambers are not invariant. They are compact pieces of the energy surface, and orbits may enter or leave them through the hemispheres.

We use the symplectic form $\omega_0=\sum_jdp_j\wedge dq_j$, and the Hamiltonian vector field is given by $\iota_{X_H}\omega_0=-dH$. All rotation numbers are computed for contractible periodic orbits, using a global trivialization of the plane field transverse to the flow. For a noncontractible retrograde orbit $P_3\subset\RP^3$, we consider its double cover $P_3^2$. Periodic orbits admit the lower semi-continuous Conley-Zehnder index $\CZ$ introduced in \cite{HWZ98}. We have
\begin{equation}\label{eq: rotation-index}
 \CZ(P)\geq3\quad\Longleftrightarrow\quad\rho(P)>1.
\end{equation}

\subsection{Global surfaces of sections below \texorpdfstring{$L_1$}{L1}}
A $2$-disk for an orbit $P\subset\RP^3$ is an immersion $u:\D\to\RP^3$ that is an embedding on the interior and whose boundary is a double cover of $P$. If the interior is transverse to the flow and every orbit outside $P$ hits it both forward and backward in time, then we say that $u$ is a rational disk-like global surface of section.

\begin{thm}\label{thm: Birkhoff conjecture}
There exists $\mu_*>0$ such that for $0<\mu<\mu_*$ and $E<L_1(\mu)$, both $\mathcal M^e_{\mu,E}$ and $\mathcal M^m_{\mu,E}$ are dynamically convex. The same index bound holds for periodic orbits in the regular part of the critical levels $E=L_1(\mu)$. For $E<L_1(\mu)$, the following statements hold in each regularized component:
\begin{enumerate}
\item[(i)] Every Birkhoff retrograde orbit $P_3$ binds a rational open book whose pages are rational disk-like global surfaces of section. More generally, this holds for every orbit transversely isotopic to a Hopf fiber.
\item[(ii)] A simple orbit $P'$ corresponding to a fixed point of the first return map forms a positive Hopf link with $P_3$. This link binds an open book whose pages are annulus-like global surfaces of section.
\end{enumerate}
\end{thm}
Here, a positive Hopf link in $\RP^3$ means that its two components lift to transverse unknots in $S^3$, each with self-linking number $-1$, and their lifts have linking number $+1$.

\subsection{Finite energy foliations through \texorpdfstring{$L_1$}{L1} and \texorpdfstring{$L_2$}{L2}}
Set
\begin{equation}\label{eq: eta}
 \eta=\left(\frac\mu3\right)^{1/3} \quad \mbox{ for every}\ 0<\mu\leq \frac12.
\end{equation}
The critical values satisfy
\begin{equation}\label{eq: lagrange-values}
 \begin{split}
 L_1(\eta)&=-\frac32-\frac92\eta^2+5\eta^3+O(\eta^4),\\
 L_2(\eta)&=-\frac32-\frac92\eta^2+7\eta^3+O(\eta^4).
 \end{split}
\end{equation}
In particular, $\frac{1}{\eta^2}(L_2(\eta)-L_1(\eta)) =2\eta+O(\eta^2)\to0$. Let $r_1>0$ and $r_2>0$ be the distances from $l_1$ and $l_2$ to the Moon, respectively. From the equilibrium equations, we obtain
\[
 \mu=\frac{r_1^3(3-3r_1+r_1^2)}{1-2r_1+r_1^2+2r_1^3-r_1^4},\qquad
 \mu=\frac{r_2^3(3+3r_2+r_2^2)}{1+2r_2+r_2^2+2r_2^3+r_2^4}.
\]
Since $\mu=3\eta^3$, we can invert the relations above to obtain
\[
 r_1(\eta)=\eta-\frac13\eta^2-\frac19\eta^3+O(\eta^4),\qquad
 r_2(\eta)=\eta+\frac13\eta^2-\frac19\eta^3+O(\eta^4).
\]
At an equilibrium, we have $p+iq=0$. Substituting $q_1=1-\mu-r_1$ and $q_1=1-\mu+r_2$ into the effective potential $U_\mu(q):= -\frac{\mu}{|q-(1-\mu)|}-\frac{1-\mu}{|q+\mu|}-\frac12|q|^2$,
we obtain \eqref{eq: lagrange-values}.

A dividing sphere $\mathcal S_i$ contains the Lyapunov orbit $P_{i,2}$ as its equator. This orbit is positive hyperbolic in its energy level. Its Conley-Zehnder index is $2$ and its rotation number is $\rho(P_{i,2})=1$. The Hamiltonian vector field is transverse to the two hemispheres in opposite directions. From \cite{LS25}, a $2-3$ foliation in a one-holed $\RP^3$ has binding $P_{i,2}\cup P_3$, one family of planes asymptotic to $P_3^2$, one rigid cylinder from $P_3^2$ to $P_{i,2}$, and the two hemispheres of $\mathcal S_i$. In a two-holed $\RP^3$, a $2-3-2$ foliation has two such families of planes and two rigid cylinders. These leaves fill the chambers, are transverse to the flow, and connect the retrograde and Lyapunov binding orbits.

Combining a $2-3$ foliation in each of the Earth and Moon chambers, we obtain a $3-2-3$ foliation for energies in between $L_1$ and $L_2$.

\begin{thm} \label{thm: FEF for mu small - I}
There exists $\mu_*>0$ such that for $0<\mu<\mu_*$ and $L_1(\mu)<E<L_2(\mu)$, the flow on $\mathcal M^1_{\mu,E}$ admits a $3-2-3$ finite energy foliation. Its binding consists of retrograde orbits $P_3^e,P_3^m$ and the Lyapunov orbit $P_{1,2}$. Each chamber contains infinitely many geometrically distinct periodic orbits and infinitely many geometrically distinct homoclinic orbits to $P_{1,2}$. If the inward stable and unstable branches in one of the chambers do not coincide, that chamber contains infinitely many transverse homoclinics to $P_{1,2}$ and a compact invariant set with positive topological entropy.
\end{thm}

Finally, we discuss the existence of finite energy foliations slightly above $L_2$, combining a $2-3$ foliation near the Earth and a $2-3-2$ foliation near the Moon, to form a $3-2-3-2$ foliation for $\mathcal M^1_{\mu,E}$.

\begin{thm}\label{thm: FEF for mu small - II}
There are $\epsilon_0,\mu_*>0$ such that, for $0<\mu<\mu_*$ and $L_2(\mu)<E<L_2(\mu)+\epsilon_0\eta^2$,
the compact region $\mathcal M^1_{\mu,E}$ admits a $3-2-3-2$ finite energy foliation. Its binding consists of the retrograde orbits $P_3^e,P_3^m$ and the Lyapunov orbits $P_{1,2},P_{2,2}$. The action of $P_{1,2}$ is greater than the action of $P_{2,2}$. Each of the two chambers $\mathcal M^e_{\mu,E}$ and $\mathcal M^m_{\mu,E}$ contains infinitely many geometrically distinct periodic orbits and infinitely many geometrically distinct homoclinic orbits to $P_{1,2}$. In addition, the compact region $\mathcal M^1_{\mu,E}$ contains infinitely many geometrically distinct homoclinic orbits to $P_{2,2}$. The homoclinics to $P_{2,2}$ may cross the first neck and enter the Earth chamber. If the inward stable and unstable branches of $P_{1,2}$ in either chamber do not coincide, that chamber contains infinitely many transverse homoclinics to $P_{1,2}$ and a compact invariant set with positive topological entropy.  If $\mathcal M^1_{\mu,E}$ contains a heteroclinic orbit between $P_{1,2}$ and $P_{2,2}$, then it contains infinitely many geometrically distinct transverse heteroclinic orbits in each direction between them.
\end{thm}

\begin{figure}[hbpt]
    \centering
    \begin{minipage}[t]{0.45\textwidth}
    \centering
    \includegraphics[scale=0.7]{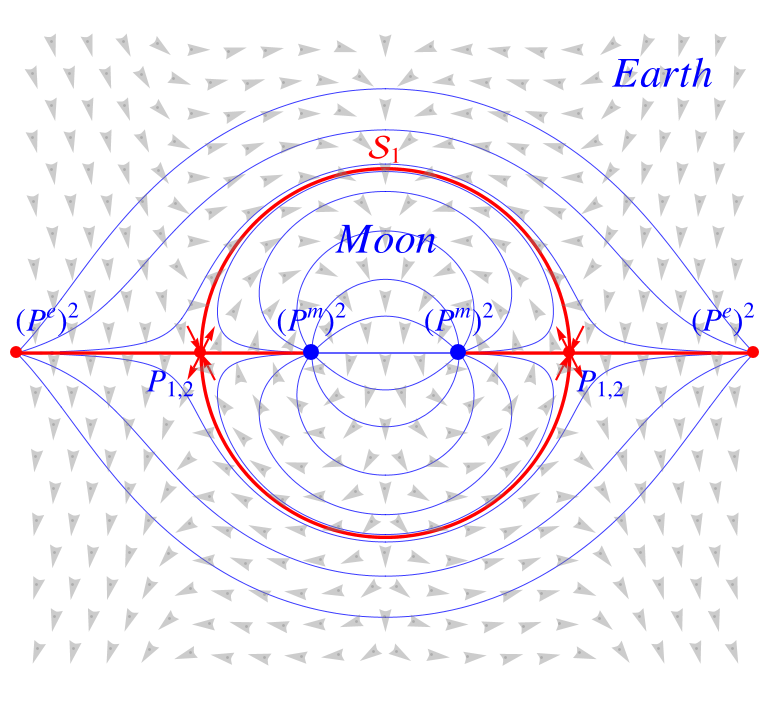}
    \end{minipage}
    \quad
    \begin{minipage}[t]{0.45\textwidth}
    \centering
    \includegraphics[scale=0.7]{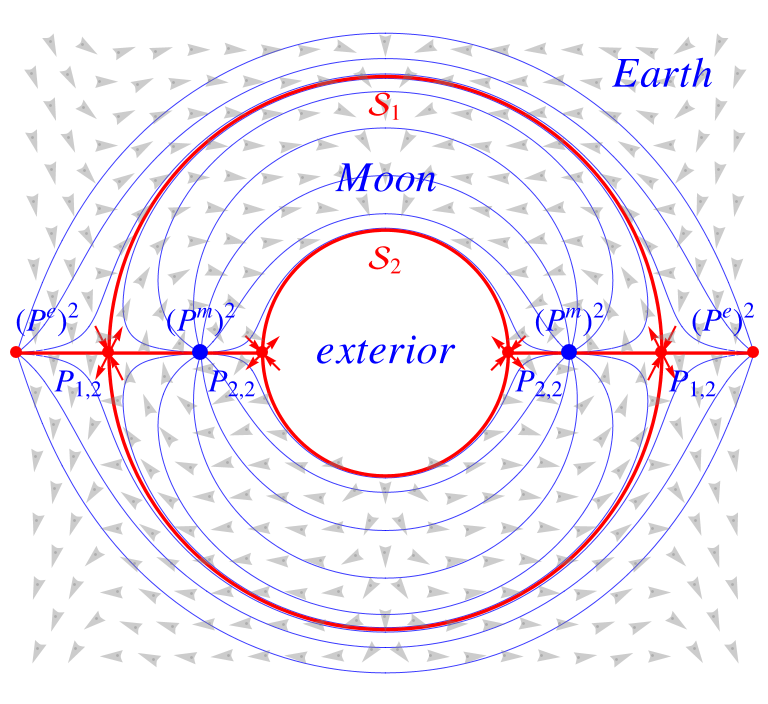}
    \end{minipage}
   \caption{Schematic finite energy foliations. Left: the $3-2-3$ foliation for $L_1<E<L_2$, when only the first neck is open. Right: the $3-2-3-2$ foliation for energies slightly above $L_2$, when both necks are open.}
    \label{fig:foliations}
\end{figure}

\subsection{Idea of the proof}
The proof has three parts. We first control the dynamics uniformly as the mass tends to zero. On the Moon side, we rescale and compare with the Hill's lunar problem. On the Earth side, we compare with the rotating Kepler problem without any rescaling. Near the saddle-centers, we prove a uniform local index estimate, ruling out the low-index periodic orbits that would obstruct the foliations.

We next construct a contact form on the whole compact region. The Liouville vector fields from the Earth and Moon sides do not match near the first neck. In fact, they have different sizes there. We join them in two steps. First, we make a specific choice at the saddle-center part, and then establish local interpolations on their overlaps. The same local construction works at the second neck. Once a single contact form and compatible dividing spheres are available, the finite energy foliation criterion from \cite{LS25}, together with the convexity result from \cite{LS26}, produces the foliations in the two chambers. They agree near the common dividing sphere and can therefore be joined.

Finally, we study the dynamical consequences of these foliations. For $P_{1,2}$, the action difference, reversibility and the infinite twist near the neck force infinitely many homoclinics. For $P_{2,2}$ we use the global foliation. We follow the circle cut out from a nearby page by its unstable manifold, preserving its $\alpha$-integral $\A(P_{2,2})$. Whenever this circle bounds a disk on the current leaf, Stokes gives that disk the same $d\alpha$-area. The circle may cross the first neck. Finiteness of the area of the rigid leaves forces it to meet a stable manifold. A returning arc is twisted infinitely near the neck and causes infinitely many intersections with the reversed arc.
For the heteroclinic assertion, the action inequality excludes local coincidence of the invariant manifolds along a connection between the two Lyapunov orbits. A nearby unstable arc becomes a spiral after passing near $P_{1,2}$. If the inward branches of $P_{1,2}$ in the exit chamber do not coincide, transverse homoclinics to $P_{1,2}$ provide stable arcs that meet this spiral in infinitely many transverse intersections. If those branches coincide, a global transition and a second passage through the neck bring the spiral back to the initial chamber, where transverse homoclinics are available and the comparison can be carried out. Reversibility gives infinitely many transverse heteroclinic orbits in both directions.

\hspace{3mm}

\noindent \textbf{AI disclosure.}
The mathematical ideas and proof strategies are due to the authors. AI tools assisted with the exposition, the revision and simplification of some proofs, and the literature review. The authors take full responsibility for the manuscript.

\section{A criterion for finite energy foliations}\label{sec: criterion}
Let
\[
 \mathcal M=L(p,q)\setminus\bigcup_{j=1}^l B_j,
 \qquad L(p,q)=S^3/\big((z_1,z_2)\sim(e^{2\pi i/p}z_1,e^{2\pi iq/p}z_2)\big),
\]
where $p,q$ are relatively prime and the $B_j$ are disjoint regular open $3$-balls. Write $\mathcal S_j=\partial B_j$. The standard contact structure on $L(p,q)$ is the quotient of the standard tight contact structure on $S^3$ determined by the Liouville form $
\alpha_0
=\frac{1}{4i}\sum_{j=1}^{2}
\left(z_j\,d\overline{z}_j-\overline{z}_j\,dz_j\right)
$.

Let $\alpha$ be the restriction of a contact form on $L(p,q)$ defining this contact structure. After a contactomorphism, it has the form $f\alpha_0$, where $f:L(p,q) \to \R^+$ is smooth. Its Reeb vector field is determined by $\alpha(R)=1$ and $\iota_Rd\alpha=0$.
For a periodic Reeb orbit $P$, write $\A(P)=\int_P\alpha$ for its action. For a contractible orbit $P=(x,T)$, consider a global symplectic trivialization of the lifted contact structure $\xi = \ker \alpha$. If $\theta(t)$ is a continuous argument of a nonzero vector under the lifted linearized transverse flow in this trivialization, define $\rho(P)=\frac{T}{2\pi}\lim_{t\to\infty}\frac{\theta(t)}t$.
The limit does not depend on the initial nonzero vector or on the trivialization of $\xi$.

A transverse $p$-unknot $P$ admits a $p$-disk $u:\D\to\mathcal M$. In the foliation criterion below, we require a $p$-disk whose characteristic foliation has one singular point, positive and nicely elliptic, as in the Bishop-family construction, see \cite{Hofer93, LS25}. A nonzero section of $u^*\ker\alpha$ gives a disjoint push-off $P^\sharp$ of the $p$-cover of $P$. Its rational self-linking number is $\slk(P)=p^{-2}(P^\sharp\cdot u)$. For the standard transverse $p$-unknot, this number is $-\frac1p$.

An adapted almost complex structure on $\R\times\mathcal M$ is $\R$-invariant, satisfies $J\partial_a=R$, preserves $\xi= \ker\alpha$, and is compatible with $d\alpha|_\xi$. Let $(\Sigma,j)$ be a connected Riemann surface. Let $\Gamma\subset \Sigma^\circ$ be a finite set of punctures. A finite energy curve is a $J$-holomorphic map $\widetilde u=(a,u):(\Sigma\setminus\Gamma,j)\longrightarrow\R\times\mathcal M$ satisfying $d\tilde u\circ j=J\circ d\tilde u$ with finite Hofer energy
\[
0<E(\widetilde u) =\sup_{\substack{\phi:\R\to[0,1]\\\phi'\geq0}}
\int_{\Sigma\setminus\Gamma}\widetilde u^*d(\phi(a)\alpha)<\infty.
\]
We call it nicely embedded if its projection $u$ is embedded. At a non-removable puncture the curve has a positive or negative end. For a nondegenerate asymptotic orbit, its asymptotic limit is unique. In the degenerate case, we use the usual statement along a subsequence.

\begin{defi}
A finite energy foliation adapted to $\alpha$ and a finite binding set $\mathcal P$ is an $\R$-invariant foliation of $\R\times\mathcal M$ by the trivial cylinders over $\mathcal P$ and nicely embedded finite energy curves, with a common energy bound. Its projection is a singular foliation of $\mathcal M$, transverse to $R$ outside the binding. In the foliations considered here, the projected leaves have genus zero and exactly one positive puncture.
\end{defi}

We use the following criterion from \cite[Theorem~1.3]{LS25}.
\begin{thm}\label{thm: criterion of FEF}
Assume that $\mathcal M$ and $\alpha$ are as above and that:
\begin{enumerate}
\item The interior contains a $p$-unknotted orbit $P_3$ with $\slk(P_3)=-\frac1p$ and $\rho(P_3^p)>1$. It admits a $p$-disk $\mathcal D\subset\mathcal M^\circ$ whose characteristic foliation has a unique positive, nicely elliptic singularity.

\item Each $\mathcal S_j\subset \partial \mathcal M$ has a hyperbolic periodic orbit $P_{2,j}$ with $\rho(P_{2,j})=1$. Its two hemispheres are projections of nicely embedded $J$-holomorphic planes asymptotic to $P_{2,j}$ from opposite directions. The Reeb vector field $R$ crosses the hemispheres in opposite directions.
\end{enumerate}
Let $S(\mathcal D,\alpha):=\int_{\D}|u^*d\alpha|.$
Suppose that there is no contractible periodic orbit $P'\subset\mathcal M\setminus(\partial\mathcal M\cup P_3)$ satisfying
\begin{equation}\label{eq: obstruction}
 \rho(P')=1,\qquad \lk(P',P_3^p)=0,
 \qquad \A(P')\leq S(\mathcal D,\alpha).
\end{equation}
Then there is an adapted $J'$, arbitrarily $C^\infty$-close to $J$ and equal to $J$ near $\partial\mathcal M$, for which $\mathcal M$ admits a finite energy foliation with binding $P_3,P_{2,1},\ldots,P_{2,l}$. The nontrivial leaves are planes and cylinders. For $l=1$ and $l=2$, its projection is a $2-3$ and a $2-3-2$ foliation, respectively.
\end{thm}

In our applications, every interior contractible orbit has a rotation number greater than $1$, thus the obstruction set is empty.
We can also use \cite[Corollary~1.4]{LS25}: a boundary orbit of maximal action has a homoclinic orbit or a heteroclinic orbit in $\mathcal M$ to another boundary orbit. If there is only one boundary sphere, the connection is homoclinic. In this one-holed case, the multiplicity conclusion of \cite{dPS2,LS25} applies.

\section{The planar circular restricted three-body problem}\label{sec: CR3BP}

In this section we introduce the regularized coordinates and study the small-mass limit. Near the Moon, the rescaled system converges to Hill's lunar problem. Near the Earth, the limit is the rotating Kepler problem. We also describe the energy levels and the two saddle-centers in the coordinates needed for the index estimates and contact construction.

Write $H_\mu=\frac12|p+iq|^2+U_\mu(q)$. Hamilton's equations are
\begin{equation}\label{equ: motion of H}
 \dot q=p+iq,\qquad \dot p=ip-\frac{\mu(q-1+\mu)}{|q-1+\mu|^3}
 -\frac{(1-\mu)(q+\mu)}{|q+\mu|^3}.
\end{equation}
Equivalently, $\ddot q_1=-2\dot q_2-\partial_{q_1}U_\mu$ and $\ddot q_2=2\dot q_1-\partial_{q_2}U_\mu$.

%\subsection{Elliptic-hyperbolic regularization}
We start with the elliptic-hyperbolic regularization for the restrict problem in general mass. It differs slightly from the one used in \cite{LS25} for $\mu$ close to $\frac12$. The two magnetic potentials differ by the differential of a smooth function, and the corresponding Hamiltonians are related by an exact symplectic shear. We use the present choice because the magnetic potential vanishes to third order at the Moon branch point, as required for the rescaling of Section~\ref{sec: smooth continuation}.

First let $\bar q=q-\frac12+\mu$ and $\bar p=p+i(1-\mu)$. Then the primaries are at $\pm\frac12$. The cotangent change
\begin{equation}\label{equ:elli-hyper. coordinates}
\bar q_1=\tfrac12\cosh x_1\cos x_2,\quad
\bar q_2=\tfrac12\sinh x_1\sin x_2,\quad \bar p=(D_x\bar q)^{-T}y
\end{equation}
is symplectic away from the primaries. If $D=\cosh^2x_1-\cos^2x_2$, its momentum matrix is
\[
 (D_x\bar q)^{-T}=\frac2D
 \begin{pmatrix}\sinh x_1\cos x_2&-\cosh x_1\sin x_2\\
 \cosh x_1\sin x_2&\sinh x_1\cos x_2\end{pmatrix}.
\]
The regularized Hamiltonian is
\begin{equation}\label{equ: hat HmuE}
 \hat H_{\mu,E}(y,x)=\frac D4(H_\mu(p,q)-E)
 =\frac12|y+\mathcal F(x)|^2+V_{\mu,E}(x),
\end{equation}
where $\mathcal F=\frac14\big(\sin x_2(\cos x_2-\cosh x_1), \sinh x_1(\cosh x_1-\cos x_2)\big)$.
The new effective potential becomes
\begin{equation}\label{eq: exact-EH-potential}
 V_{\mu,E}=-\frac E4D-\frac12\cosh x_1
 +\frac{1-2\mu}{2}\cos x_2
 -\frac D8\left|\bar q(x)+\frac12-\mu\right|^2.
\end{equation}
In particular, the regularization is smooth at collisions. The involution $(y,x)\mapsto(-y,-x)$ gives the double cover described in the introduction.
The form $\mathcal F\cdot dx$ is a magnetic potential. Its exterior derivative is the magnetic two-form. To compare with the regularization in \cite{LS25}, put $\mathcal F_{\rm old}=\frac18(\sin 2x_2,\sinh 2x_1), g(x)=\frac14\sinh x_1\sin x_2$.
Then $\mathcal F=\mathcal F_{\rm old}-\nabla g$. Thus the two formulas define the same magnetic two-form and are related by the exact shear $y\mapsto y+\nabla g$.

\subsection{Hill's lunar problem}\label{sec: Hill's problem}
In coordinates around the Moon, replace $(p,q)$ by $(p-i(1-\mu),q+1-\mu)$ and add $\frac12(1-\mu)^2$ to $H_\mu$. The resulting Hamiltonian is $$\tilde H_\mu(p,q)=\frac12|p+iq|^2-\frac\mu{|q|}-\frac12|q|^2 -(1-\mu)\big(\frac1{|1+q|}+q_1\big).$$
With $\eta=(\frac\mu3)^{1/3}$, set $$H^{\mathrm{Hill}}_\mu(p,q)=\eta^{-2} \big(\tilde H_\mu(\eta p,\eta q)+1-\mu\big).$$
On every fixed compact set disjoint from \(q=0\), the Earth \(q=-\frac1\eta\) escapes to infinity, and Taylor expansion at \(\eta=0\) gives
\begin{equation}\label{equ: Hamiltonian Lunar}
 H^{\mathrm{Hill}}_\mu\longrightarrow
 H(p,q)=\frac12|p+iq|^2-\frac3{|q|}-\frac32q_1^2
 \quad\text{in }C^\infty.
\end{equation}
The corresponding energy is exactly
\begin{equation}\label{eq: physical-Hill-energy}
 \bar E=\frac1{\eta^2}\left(E+\frac32-2\mu+\frac{\mu^2}{2}\right).
\end{equation}
The two critical configurations are \(q=(\pm1,0)\), with corresponding phase points \(p=-iq\), and the critical value is \(\bar E_c=-\frac92\). These points correspond to the Hill limits of $l_1$ and $l_2$.

Apply Levi-Civita coordinates $q=v^2$, $p=\frac{u}{2\bar v}$, and set $\hat K_0=4|v|^2(H+c)$, where $c=-\bar E>0$. The change $(u,v)=(c^{3/4}y,c^{1/4}x)$ gives
\begin{equation}\label{equ: Ham K}
 \begin{split}
 K(y,x)&=c^{-3/2}\big(\hat K_0(c^{3/4}y,c^{1/4}x)+12\big)\\
 &=\frac12(y_1-2|x|^2x_2)^2+\frac12(y_2+2|x|^2x_1)^2
 +2|x|^2\big(2-3(x_1^2-x_2^2)^2\big).
 \end{split}
\end{equation}
Thus \(H=-c\) corresponds to \(K=12c^{-3/2}\). The critical value of \(K\) is \(\frac{8\sqrt2}{9}\).

Let $G_0(x)=\frac23(x_1^2-x_2^2)x_1x_2$. The shear $(y,x)\mapsto(y+\nabla G_0(x),x)$ is symplectic. We then apply the constant linear substitution $y\mapsto \frac83y$. This change is conformally symplectic with a positive constant factor, so it preserves characteristic trajectories up to a positive time change. With a corresponding positive rescaling of the Hamiltonian, we obtain
\begin{equation}\label{equ: hat K0}
 \begin{split}
 \hat K(y,x)&=\frac9{64}K\left(\frac83y+\nabla G_0(x),x\right)\\
 &=\frac12(y_1-x_2^3)^2+\frac12(y_2+x_1^3)^2
 +\frac9{32}|x|^2\big(2-3(x_1^2-x_2^2)^2\big).
 \end{split}
\end{equation}
The Hamiltonian \(\hat K\) in \eqref{equ: hat K0} is exactly the Hill Hamiltonian studied in \cite{LS26}, with the same normalization. We will use several results from that paper for this Hamiltonian.

Put
\begin{equation}\label{eq: Hill-saddles}
 a_{\rm H}=\frac{2^{1/4}}{\sqrt3},\quad \h_c=\frac1{4\sqrt2},\quad
 S_{1,\pm}=(\pm a_{\rm H}^3,0,0,\pm a_{\rm H}),\quad
 S_{2,\pm}=(0,\mp a_{\rm H}^3,\pm a_{\rm H},0).
\end{equation}
The first pair of points corresponds to the Earth-facing Hill equilibrium. All four points are saddle-centers on $\hat K^{-1}(\h_c)$.

We use the convexity result of \cite{LS26} in the rescaled coordinates above.

\begin{prop}[\cite{LS26}]\label{prop: Hill convexity}
For $c>\frac92$, the bounded regularized component of $\hat K^{-1}(\frac{27}{16}c^{-3/2})$ is strictly convex. On the bounded critical component at $c=\frac92$, its second fundamental form is positive at every regular point.
\end{prop}

At the critical value $c=\frac92$, the bounded component is convex and has positive second fundamental form at every regular point, as used below.

\subsection{Smooth continuation from Hill to the \texorpdfstring{$\mu$}{mu}-small restricted problem}\label{sec: smooth continuation}
We establish smooth convergence of the regularized small-mass Hamiltonians to Hill's lunar problem. We also follow the two saddle-centers separately, since the difference between $L_1$ and $L_2$ appears only at higher order.

Let $r=r_1$ be the distance from $l_1$ to the Moon. The equilibrium equation gives a real-analytic function near $r=0$
\begin{equation}\label{eq: mu-r}
 \mu(r)=\frac{r^3(3-3r+r^2)}{1-2r+r^2+2r^3-r^4}.
\end{equation}
It is strictly increasing for small positive $r$, and
$\eta(r)=\left(\frac{\mu(r)}{3}\right)^{1/3}=r+\frac{r^2}{3}+O(r^3)$
is also real-analytic near $r=0$. We parametrize energy by
\begin{equation}\label{equ: bar E and c}
 E(r,c)=L_1(\mu(r))+(\frac92-c)\eta(r)^2,
\end{equation}
which is real-analytic in $(r,c)$ near $(0,\frac92)$. In this formula, $c$ is a rescaled energy parameter. In particular,
\begin{equation}\label{eq: E-expansion}
 E(r,c)=-\frac32-cr^2+\left(5-\frac{2c}{3}\right)r^3+O(r^4),
\end{equation}
uniformly for $c$ in compact subsets of $(0,\infty)$. Equation~\eqref{eq: physical-Hill-energy} becomes $\bar E=-c+O(r)$.

\begin{figure}[htbp]
\centering
\includegraphics[width=\textwidth]{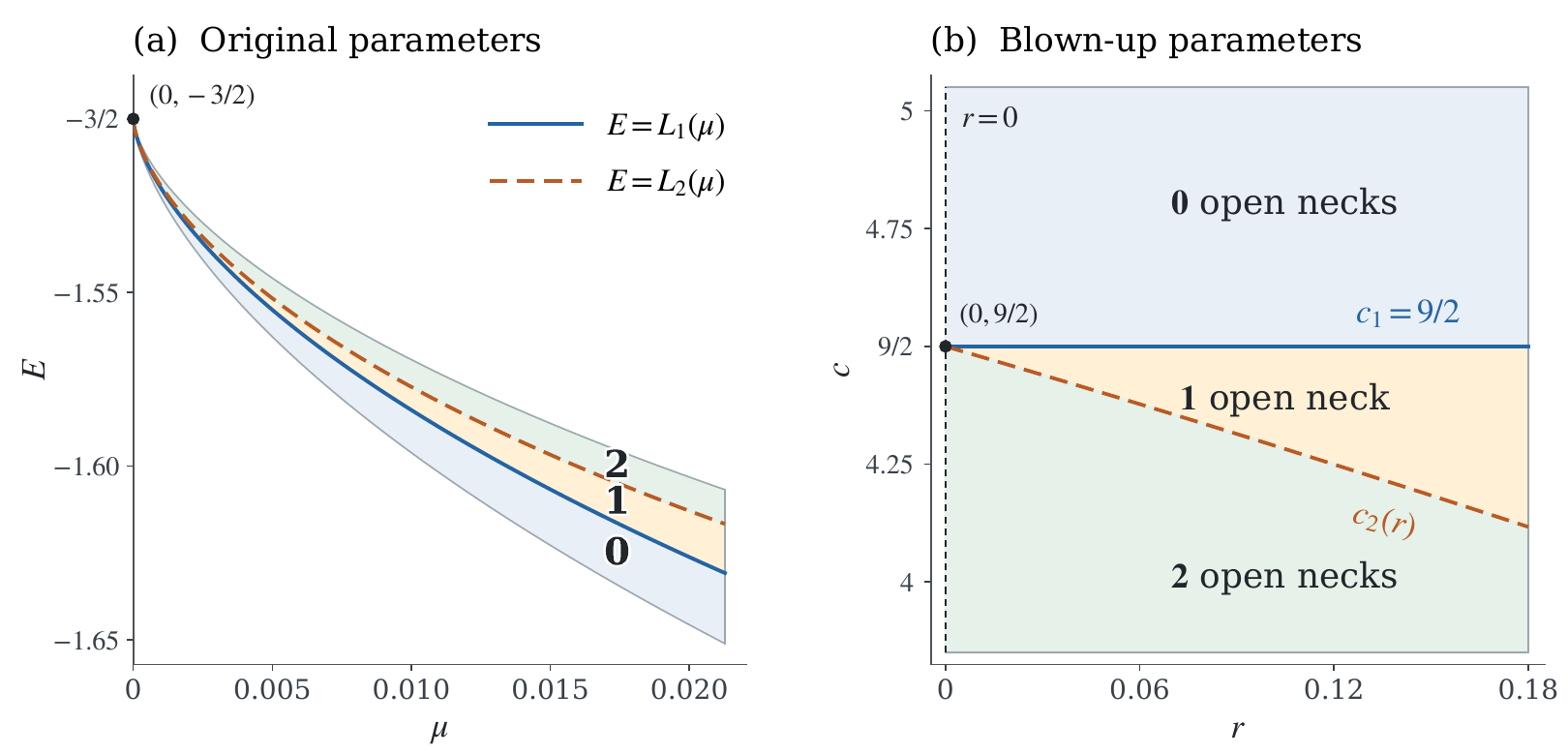}
\caption[Critical curves under the parameter blow-up]{Critical curves
in the original parameters $(\mu,E)$ (left) and the blown-up parameters $(r,c)$ (right).
The curves $E=L_1(\mu)$ and $E=L_2(\mu)$ become $c=c_1=\frac92$ and $c=c_2(r)=\frac92-\eta(r)^{-2} \bigl(L_2(\mu(r))-L_1(\mu(r))\bigr)$, respectively. They meet at $(0,\frac92)$ with distinct tangents: $c_2(r)=\frac92-2r+O(r^2)$. The entire edge $r=0$ maps to $(\mu,E)=(0,-\frac32)$.}
\label{fig:mu-small-parameter-blowup}
\end{figure}

For \(r>0\) small, the parameters \((r,c)\) are equivalent to the parameters \((\mu,E)\). The function \(r\mapsto\mu(r)\) is strictly increasing, and \(E(r,c)\) is affine in \(c\). At \(r=0\), this parametrization blows up the single point \((\mu,E)=(0,-\frac32)\) into the family \(\{(0,c):c>0\}\), which corresponds to the energy on the Hill scale. Thus \(c\) measures the energy relative to \(L_1(\mu)\) on the natural Hill scale \(\eta^2\). The curves \(c=\mathrm{const}\) parametrize the nearby energy levels, with \(c=\frac92\) corresponding exactly to \(E=L_1(\mu)\). See Figure \ref{fig:mu-small-parameter-blowup}.

Near the Moon branch point $x=0$, define $\hat G_0(x)=\frac23\left(\sinh^2\frac{x_1}{2}-\sin^2\frac{x_2}{2}\right) \sinh\frac{x_1}{2}\sin\frac{x_2}{2}$.
This is a local generating function. We multiply it by an even cutoff function supported in this neighborhood and equal to one in a smaller neighborhood of the Moon.
Set $K_{\mu,E}(y,x)=\hat H_{\mu,E}(y+\nabla\hat G_0(x),x)$.
The new magnetic potential satisfies
\begin{equation}\label{eq: sheared-magnetic-expansion}
 \mathcal F+\nabla\hat G_0
 =\big(-\frac{x_2^3}{6},\frac{x_1^3}{6}\big)+O(|x|^5).
\end{equation}
Now let
\begin{equation}\label{equ: hat Kr1}
 \begin{split}
 \phi_{r,c}(y,x)&=\left(\frac43(r^2c)^{3/4}y,\,2(r^2c)^{1/4}x\right),\\
 \hat K_{r,c}&=\frac9{16(r^2c)^{3/2}}
 \left(K_{\mu(r),E(r,c)}\circ\phi_{r,c}+3r^3\right),\qquad
 \h(c)=\frac{27}{16c^{3/2}}.
 \end{split}
\end{equation}
The zero level of $\hat H_{\mu(r),E(r,c)}$ corresponds to $\hat K_{r,c}^{-1}(\h(c))$. Also,
\begin{equation}\label{eq: conformal-factor}
 \phi_{r,c}^*\omega_0=\frac83r^2c\,\omega_0.
\end{equation}
Hence, the change is conformally symplectic with a positive constant factor and preserves the Hamiltonian trajectories up to a positive time change. Since $\phi_{r,c}$ contains $r^{1/2}, r^{3/2}$ etc, it is not smooth at $r=0$. However, after substitution and cancellation, $\hat K_{r,c}$ depends smoothly on $(r,c)$ near $(0,\frac92)$ and converges to the regularized Hill Hamiltonian $\hat K$ in $C^\infty_{\rm loc}$ as $(r,c) \to (0,\frac92)$.

\begin{prop}\label{prop: smooth continuation}
The family $\hat K_{r,c}$ extends smoothly to $r=0$ with $\hat K_{0,c}=\hat K$. More precisely, for a compact interval $C\subset(0,\infty)$, a compact $B\subset\R^4$, and every $k\geq0$, $\|\hat K_{r,c}-\hat K\|_{C^k(B)}\leq C_{B,C,k}r, \forall c\in C$.
Regular subsets of the corresponding energy levels converge in smooth local graphs.
\end{prop}

\begin{proof}
Since \(\frac{\mu(r)}3=r^3a(r)\), where \(a\) is real-analytic and
\(a(0)=1\), we have $\eta(r)=ra(r)^{1/3}$. Thus \(\eta(r)\), \(L_1(\mu(r))\) and \(E(r,c)\) are smooth, in fact real-analytic in \(r\) near \(r=0\). In particular, $E(r,c)=-\frac32-cr^2+O(r^3)$, uniformly for \(c\) in compact subsets of \((0,\infty)\).

We prove smoothness of the rescaled Hamiltonian at \(r=0\). Let $s=\sqrt r$.
Although the rescaling \(\phi_{r,c}\) contains half-integer powers of \(r\),
we will see that these powers disappear from the Hamiltonian.

Set $A(x):=\mathcal F(x)+\nabla\hat G_0(x)$.
The function \(A\) is real analytic and odd under \(x\mapsto-x\). By
\eqref{eq: sheared-magnetic-expansion}, $A(x) =\frac16\left(-x_2^3,x_1^3\right) +O(|x|^5)$. Since only odd powers occur, we can write $A(2sc^{1/4}x) = \frac43s^3c^{3/4}(-x_2^3,x_1^3) +s^5A_1(s^2,c,x)$, where \(A_1\) is smooth in \((s^2,c,x)\). Hence
\[
\begin{split}
&\frac43s^3c^{3/4}y+A(2sc^{1/4}x) =
\frac43s^3c^{3/4}
\big(y_1-x_2^3,\,y_2+x_1^3\big)
+s^5A_1(s^2,c,x).
\end{split}
\]
It follows that the kinetic part satisfies
\[
\begin{split}
\frac12\left|
\frac43s^3c^{3/4}y+A(2sc^{1/4}x)
\right|^2
&=
\frac89s^6c^{3/2}
\left[
(y_1-x_2^3)^2+(y_2+x_1^3)^2
\right]\\
&\quad+s^8R_1(s^2,c,y,x),
\end{split}
\]
with \(R_1\) smooth in $(s^2,c,y,x)$.

We next consider the potential. The expression \eqref{eq: exact-EH-potential} of $V_{\mu,E}(x)$ is real-analytic in \((\mu,E,x)\) near
\((0,-\frac32,0)\), and it is even under \(x\mapsto-x\). Its Taylor expansion
at \(x=0\), together with the expansions of \(\mu(r)\) and \(E(r,c)\), gives
\begin{equation}\label{potent}
\begin{split}
V_{\mu(r),E(r,c)}(x)
={}&-\mu(r)
+\frac{cr^2}{4}|x|^2
-\frac3{128}|x|^2(x_1^2-x_2^2)^2\\
&\quad+r^3Q_2(r,c,x)
+r^2R_4(r,c,x)
+R_8(r,c,x),
\end{split}
\end{equation}
where \(Q_2\) is quadratic in \(x\), $R_4(r,c,x)=O(|x|^4), R_8(r,c,x)=O(|x|^8)$, and all these functions are smooth. Since the
potential is even in \(x\), the Taylor expansions of \(R_4\) and \(R_8\)
contain only even total degrees.

Now substitute \(r=s^2\) and \(x\mapsto2sc^{1/4}x\). Since $-\mu(s^2)+3s^6=O(s^8)$,
the constant term in \eqref{potent} contributes only order \(s^8\). The quadratic term gives $\frac{cs^4}{4}\, |2sc^{1/4}x|^2 =s^6c^{3/2}|x|^2$,
while the sextic term gives
\[
-\frac3{128}
\left|2sc^{1/4}x\right|^2
\left(
(2sc^{1/4}x_1)^2-(2sc^{1/4}x_2)^2
\right)^2
=
-\frac32s^6c^{3/2}
|x|^2(x_1^2-x_2^2)^2.
\]
Moreover, by the
evenness in \(x\), after the substitution all remaining terms are \(s^8\) times a smooth
function of \(s^2\), rather than only a smooth function of \(s\).
Therefore
\[
\begin{split}
&V_{\mu(s^2),E(s^2,c)}
(2sc^{1/4}x)+3s^6\\
&\qquad =
s^6c^{3/2}
\left[
|x|^2-\frac32|x|^2(x_1^2-x_2^2)^2
\right]
+s^8R_3(s^2,c,x),
\end{split}
\]
where \(R_3\) is smooth.

Combining the kinetic and potential terms, we obtain
\begin{align*}
&K_{\mu(s^2),E(s^2,c)}
\left(\frac43s^3c^{3/4}y,\,
2sc^{1/4}x\right)+3s^6\\
&\quad =s^6c^{3/2}\Bigg[\frac89
\left((y_1-x_2^3)^2+(y_2+x_1^3)^2
\right)+|x|^2-\frac32|x|^2(x_1^2-x_2^2)^2\Bigg]\\
&\qquad\quad +s^8R(s^2,c,y,x),
\end{align*}
for a smooth function \(R\).

From the definition of $\hat K_{r,c}$ we conclude that
\[
\hat K_{r,c}(y,x)=
\frac12(y_1-x_2^3)^2+\frac12(y_2+x_1^3)^2+\frac9{32}|x|^2
\left(2-3(x_1^2-x_2^2)^2\right)
+r\,\widetilde R(r,c,y,x),
\]
where \(\widetilde R\) is smooth in \((r,c,y,x)\) near
\(r=0\), uniformly for \(c\) in compact subsets of \((0,\infty)\).
The first three terms are precisely \(\hat K\). Hence $\hat K_{r,c}=\hat K+r\,\widetilde R(r,c,y,x)$.

This proves that \(\hat K_{r,c}\) extends smoothly to \(r=0\)
in coordinates \((r,c,y,x)\), with $\hat K_{0,c}=\hat K$. The same identity also gives, for every compact \(C\subset(0,\infty)\), every compact \(B\subset\mathbb R^4\), and every \(k\geq0\), $\|\hat K_{r,c}-\hat K\|_{C^k(B)} \leq C_{B,C,k}\,r, c\in C$. Finally, if \(d\hat K\neq0\), the implicit function theorem and
the smooth dependence on \((r,c)\) imply that the corresponding regular energy levels converge in smooth local graphs.
\end{proof}

\begin{lem}\label{lem: moving-critical-values}
For \((r,c)\) sufficiently close to \((0,\frac92)\), each critical point
\(S_{i,\pm}\) of \(\hat K\) admits a unique smooth continuation $S_{i,\pm}(r,c)$ as a critical point of \(\hat K_{r,c}\), \(i=1,2\). The antipodal involution \((y,x)\mapsto(-y,-x)\) exchanges
\(S_{i,+}(r,c)\) and \(S_{i,-}(r,c)\). Hence, they have the same critical value, which we denote by $k_i(r,c):= \hat K_{r,c}\big(S_{i,\pm}(r,c)\big)$.
Set $\delta_i(r,c):=\h(c)-k_i(r,c)$,
and define $c_1(r):=\frac92, c_2(r):=\frac92-\eta(r)^{-2}(L_2(\mu(r))-L_1(\mu(r)))$.
Then $E(r,c_i(r))=L_i(\mu(r))$,
and there exist smooth positive functions \(a_i(r,c)\) such that
\begin{equation}\label{eq: actual-neck-excess}
\delta_i(r,c)
=
a_i(r,c)\big(c_i(r)-c\big).
\end{equation}
In particular, $\operatorname{sign}\delta_i(r,c) = \operatorname{sign}\big(E(r,c)-L_i(\mu(r))\big)$.
\end{lem}

\begin{proof}
By Proposition~\ref{prop: smooth continuation}, the family
\(\hat K_{r,c}\) depends smoothly on \((r,c)\) near \((0,\frac92)\). Since \(\nabla^2\hat K\) is nonsingular at \(S_{i,\pm}\), the implicit function theorem gives the unique smooth continuations
\(S_{i,\pm}(r,c)\). The regularized Hamiltonian is invariant under the  involution \((y,x)\mapsto(-y,-x)\). Hence, this involution exchanges
\(S_{i,+}(r,c)\) and \(S_{i,-}(r,c)\), and the two points have the same critical value \(k_i(r,c)\).

By definition of \(E(r,c)\), we have $E(r,c_1(r))=E(r,\frac92) =L_1(\mu(r))$. Likewise, the definition of \(c_2(r)\) gives $E(r,c_2(r))=L_2(\mu(r))$.
Therefore, at \(c=c_i(r)\), the energy level contains the
equilibrium \(l_i(\mu(r))\). Its two lifts under the regularization are precisely \(S_{i,\pm}(r,c_i(r))\). Hence $\delta_i(r,c_i(r)) = \h(c_i(r))-k_i(r,c_i(r)) =0$.

At \(r=0\), all four limiting critical points lie on the critical Hill level, so $k_i(0,c)=\h_c$
near \(c=\frac92\). Therefore, $\partial_c\delta_i(0,\frac92) = \h'(\frac92) = -\frac{1}{12\sqrt2}<0$. After shrinking the parameter neighborhood, \(\partial_c\delta_i\)
remains negative. Since \(\delta_i(r,c_i(r))=0\), division in the \(c\)-variable gives $\delta_i(r,c) = a_i(r,c)\big(c_i(r)-c\big)$, where \(a_i\) is smooth and positive.

Finally, $E(r,c)-L_i(\mu(r)) = \eta(r)^2\big(c_i(r)-c\big)$,
so \(\delta_i(r,c)\) and \(E(r,c)-L_i(\mu(r))\) have the same sign.
\end{proof}

Notice that for $c\neq c_i(r)$, the point $S_{i,\pm}(r,c)$ is a critical point of the regularized Hamiltonian $\hat K_{r,c}$, not necessarily the image of an equilibrium on $H_{\mu(r)}=E(r,c)$.

\subsection{The rotating Kepler problem}\label{subsec: Rotating Kepler problem}
The dynamical convexity of the rotating Kepler problem below the critical energy and the Conley-Zehnder indices of its periodic orbits in that range were established in \cite{AlbersFishFrauenfelderVanKoert13}. A convex symplectic embedding, together with a perturbative extension near the heavy primary away from collisions, was obtained in \cite{FrauenfelderVanKoertZhao22}. We use the following explicit description to obtain uniform estimates near the critical energy.

For $\mu=0$, the Hamiltonian becomes $H_0(p,q)=\frac12|p|^2-\frac1{|q|}+q_1p_2-q_2p_1$.
Its critical value is $-\frac32$, and its critical set is the circle $\mathcal C=\{|q|=1,p=-iq\}$. We recall the coordinates used by McGehee \cite{McGehee1969}. Let $(p,q)=(\frac w{\bar z},2z^2), K_{0,E}=\frac12|w|^2-\frac12+i(z\bar w-\bar zw)|z|^2-E|z|^2$.
On the bounded regularized component, set
\[
 h^2=-2E+2(q_1p_2-q_2p_1),\quad
 \zeta_1=w+ihz,\quad \zeta_2=\bar w+ih\bar z,
 \quad \psi=\frac{z\bar w+\bar zw}{2h^2},
\]
and $\xi_1=\zeta_1e^{i\psi}$, $\xi_2=\zeta_2e^{-i\psi}$. The coordinates $(\xi_1,\xi_2)$ are not symplectic. They identify the regularized component for $E<-\frac32$ with $\mathbb S_2=\{|\xi_1|^2+|\xi_2|^2=2\}$ and put the flow, with a positive time change, in the form
\begin{equation}\label{equ: motion on the round 3-sphere}
\dot\xi_1=i(h+h^{-2})\xi_1,\qquad
 \dot\xi_2=i(h-h^{-2})\xi_2.
\end{equation}
The relation between $h$ and the invariant tori is
\begin{equation}\label{eq: h-torus}
 |\xi_1|^2=1+(E+\frac12h^2)h.
\end{equation}
For $E=-\frac32$, this parametrization is used away from $\mathcal C$ and extends to it continuously.

The retrograde and direct circles are $\xi_2=0$ and $\xi_1=0$. For their closed lifts to $S^3$, the transverse rotation numbers are $\rho_r=\frac{2h_r^3}{h_r^3+1}$, $ \rho_d=\frac{2h_d^3}{h_d^3-1}$, where $-E=\frac12h_r^2-\frac1{h_r}=\frac12h_d^2+\frac1{h_d}$. For $E\leq-\frac32$, we have $h_r\geq2$. For $E<-\frac32$, we have $h_d>1$. Thus $\rho_r>1$ and $\rho_d>2$. At $E=-\frac32$, the retrograde value is $\frac{16}9$, and its orbit is nondegenerate. The implicit function theorem therefore gives a smooth family $P^e_{3,\mu,E}$ for $(\mu,E)$ in a fixed neighborhood of $(0,-\frac32)$.

At $E=-\frac32$, relation \eqref{eq: h-torus} becomes $|\xi_1|^2=\frac12(h-1)^2(h+2)$.
On the double cover, the return map to $\arg\xi_1=0$ has rotation
number $\rho(h)=\frac{h^3-1}{h^3+1}$. Choose a sufficiently small positive Diophantine number $\rho_0$, and define
$h_0=(\frac{1+\rho_0}{1-\rho_0})^{1/3}$ and $\chi_0=(h_0-1)\sqrt{\frac{1}{2}(h_0+2)}.$
Then the torus $T_0=\{|\xi_1|=\chi_0\}$ in the critical level has
rotation number $\rho_0$. As $\rho_0$ tends to zero, $h_0$ tends
to $1$ and $\chi_0$ tends to zero. We can therefore choose $\chi_0>0$ as small as needed. The two-torus $T_0$ is disjoint from the critical circle $\mathcal C$, so it stays a positive distance from the Moon branch point in the regularized coordinates. We work in a fixed neighborhood of $T_0$. For $\mu=0$ and nearby energies, keeping $h=h_0$ gives invariant tori with the same rotation number $\rho_0$. Since $\rho'(h)=\frac{6h^2}{(h^3+1)^2}>0$, the return map to $\arg \xi_1=0$ has a nonzero twist. The nearby Hamiltonian return maps are
exact area-preserving maps and depend smoothly on $(\mu,E)$.
Hence, the usual KAM theory, see for instance Moser's invariant curve theorem~\cite{Moser62}, gives an invariant curve with rotation number $\rho_0$ for every $(\mu,E)$
sufficiently close to $(0,-\frac32)$. This invariant curve corresponds to an invariant torus $T_{\mu,E}$.

\subsection{The low-energy dynamical convexity }\label{sec: the lower-energy range}
The Hill rescaling controls energies on the $\eta^2$ scale near the first critical value. For $c\to\infty$, we use a separate Kepler estimate.

\begin{lem}\label{lem: Kepler-tail}
There are constants $C>\frac92$, $K>0$ and $\mu_0>0$ with the following properties. For $0<\mu<\mu_0$, the regularized Moon component is dynamically convex when $c\geq C$, and the regularized Earth component is dynamically convex when $E\leq-K$.
\end{lem}

\begin{proof}
We use the same rescaling near each primary. If its mass is $m$, a translation of position and momentum puts the Hamiltonian in the form
\[
H=H_*+\frac12|p|^2-\frac{m}{|q|}
+q_1p_2-q_2p_1+W(q), \quad H_*\in \R\ \ \text{and}\ \ W(0)=dW(0)=0.
\]
Near both primaries, the $C^2$-norm of $W$ on a fixed small disk is bounded uniformly for every sufficiently small $\mu$.

Fix $E<H_*$ and set $\kappa=H_*-E>0$. Under the rescaling $q=\frac{m}{\kappa}Q$ and $p=\sqrt{\kappa}\,P$,
the equation $\frac1\kappa(H-E)=0$ becomes
\[
 \frac12|P|^2-\frac1{|Q|}+1
 +\frac{m}{\kappa^{3/2}}(Q_1P_2-Q_2P_1)
 +\frac1\kappa W\!\left(\frac{m}{\kappa}Q\right)=0.
\]
Now use the Levi-Civita coordinates $Q=v^2,P=\frac{u}{2\bar v}$ and multiply by $4|v|^2$. The resulting Hamiltonian extends smoothly
across $v=0$ and takes the form
$\mathcal K_{m,\kappa}(u,v) =\frac12|u|^2+4|v|^2-4+\mathcal R_{m,\kappa}(u,v),$
where the remainder is
\[
 \mathcal R_{m,\kappa}(u,v)
 =\frac{2m}{\kappa^{3/2}}|v|^2(v_1u_2-v_2u_1)
  +\frac{4|v|^2}{\kappa}
    W\!\left(\frac{m}{\kappa}v^2\right).
\]
Without the perturbation, the zero level is the ellipsoid
$\Sigma_0=\left\{(u,v):\frac12|u|^2+4|v|^2=4\right\}.$

Choose a fixed closed ball $B$ centered at the origin and containing $\Sigma_0$ in its interior. Since $W$ vanishes to second order at the
origin, we obtain $\|\mathcal R_{m,\kappa}\|_{C^2(B)} \leq A\left(\frac{m}{\kappa^{3/2}} +\frac{m^2}{\kappa^3}\right)$,
provided $\frac m\kappa$ is small enough such that $\frac m\kappa v^2$ stays in the disk where $W$ is controlled. The constant $A$ is independent of $\mu$. Thus, when $m\kappa^{-3/2}$ and $\frac m\kappa$ are small, $\mathcal K_{m,\kappa}$ has positive definite Hessian on $B$ and is positive on $\partial B$. Its zero set in $B$ bounds a smooth strictly convex domain.

We consider both sides separately. For the Moon, $m=\mu=3\eta^3$ and $\kappa=c\eta^2+O(\mu)$. Hence, for $c\geq C$ and sufficiently small $\mu$, there is a uniform constant $A_1$ such that $\frac{m}{\kappa^{3/2}}\leq A_1c^{-3/2}, \frac{m}{\kappa}\leq A_1\frac{\eta}{c}$.
For the Earth, $m=1-\mu\leq1$ and $H_*$ remains bounded as $\mu\to0$. Thus $\kappa\to+\infty$ uniformly as $E\to-\infty$,
and both quantities tend to zero.

We first choose $C$ and $K$ large, and then choose $\mu_0>0$ small.
For every $0<\mu<\mu_0$, the Moon component with $c\geq C$ and
the Earth component with $E\leq-K$ are therefore strictly convex
on the Levi-Civita double cover. The convexity theorem of
Hofer--Wysocki--Zehnder~\cite{HWZ98} gives the required dynamical
convexity.
\end{proof}
\section{Uniform index estimates}\label{sec: index-estimates}
The foliation criterion Theorem~\ref{thm: criterion of FEF} requires us to exclude interior contractible periodic orbits with rotation number equal to $1$. We shall prove a stronger statement: every such orbit, geometrically distinct from the Lyapunov orbits, has rotation number strictly greater than $1$. The difficulty is to make this assertion uniform as $\mu\to0$, when the necks approach one another on the Hill scale.

There are two different reasons why an orbit must have large index. On compact regular parts of the limiting energy levels we can find a transverse frame in which the linearized flow always rotates in the positive direction. Near a saddle-center, on the other hand, an orbit that comes close to the equilibrium spends a long time there, while its elliptic component keeps rotating.

Recall that the transverse linearized flow acts on the two-dimensional quotient of the tangent space by the flow direction. In a symplectic trivialization, each nonzero transverse solution determines a time-dependent differentiable argument. We call the frame positive when this angle has strictly positive derivative along every solution. On a compact set, a positive lower bound on the angular speed implies a lower bound on the total rotation in terms of the time spent there. The argument also uses the positive frame to bound the periods of low-index orbits. Such a period bound allows us to pass to a limiting periodic orbit, where dynamical convexity gives the contradiction.

\subsection{A positive frame for the Rotating Kepler Problem}\label{sec_positive_frame}
Write $\xi_j=u_j+iv_j,j=1,2$, and denote by $X_E$ and $\phi_E^t$ the vector field and flow in \eqref{equ: motion on the round 3-sphere}. On $\mathbb S_2$, consider
\begin{equation}\label{eq: RK-frame}
 \begin{split}
 X_1&=v_2\partial_{u_1}-v_1\partial_{u_2}
       +u_2\partial_{v_1}-u_1\partial_{v_2},\\
 X_2&=-u_2\partial_{u_1}+u_1\partial_{u_2}
       +v_2\partial_{v_1}-v_1\partial_{v_2}.
 \end{split}
\end{equation}
The point of this particular frame is that it is defined on the whole three-sphere, including the two circular orbits. The two vector fields span the standard contact planes. For $\lambda_{\mathrm{std}}=\frac12\sum_j(u_j\,dv_j-v_j\,du_j)$, we have $\lambda_{\mathrm{std}}(X_E) =\frac{2E+3h^2}{2h}>0$
on the bounded component if $E<-\frac32$. Thus $(X_1,X_2)$ is a global transverse frame.

We compute the angular motion in this frame. Fix $E<-\frac32$ and initial phases $\theta_{1,0},\theta_{2,0}$. For $0<\vartheta<\frac{\pi}2$, let $h_\vartheta>1$ be the branch determined by
\[
\cos(2\vartheta)=\left(E+\frac12h_\vartheta^2\right)h_\vartheta,\qquad \omega_1(\vartheta)=h_\vartheta+h_\vartheta^{-2},\quad
\omega_2(\vartheta)=h_\vartheta-h_\vartheta^{-2}.
\]
The curves
\[
 \xi_\vartheta(t)=\left(
 \sqrt2\cos\vartheta\,e^{i(\omega_1(\vartheta)t+\theta_{1,0})},
 \sqrt2\sin\vartheta\,e^{i(\omega_2(\vartheta)t+\theta_{2,0})}
 \right)
\]
are a family of solutions at the same energy. Choose $\vartheta_0\in(0,\frac{\pi}2)$ and set $\xi(t)=\xi_{\vartheta_0}(t)$ and $h=h_{\vartheta_0}$. The derivative
\begin{equation}\label{eq: RK-variational-vector}
V(t):=\left.\partial_\vartheta\xi_\vartheta(t)\right|_{\vartheta=\vartheta_0}
 =D\phi_E^t(\xi(0))V(0)
 \in T_{\xi(t)}\mathbb S_2
\end{equation}
is a solution of the linearized equation along the orbit $\xi(t)$. It is important to differentiate $h_\vartheta$ as well: the family is taken at fixed energy, not at fixed $h$. Indeed, $\partial_\vartheta h_\vartheta =-\frac{2\sin(2\vartheta)}{E+3h^2/2}$.

To obtain its transverse component, decompose $V(t)=a_0(t)X_E(\xi(t))+a_1(t)X_1(\xi(t)) +a_2(t)X_2(\xi(t))$.
Writing
\[
\sigma(t)=2ht+\theta_{1,0}+\theta_{2,0},\qquad
b=\frac{12\sin^2(2\vartheta_0)}
{(2E+3h^2)(h^3+\cos(2\vartheta_0))}\geq0,
\]
direct differentiation gives $a_1(t)+ia_2(t)=e^{i\sigma(t)}(bt+i)$.
Varying the initial phases in the direction $(\tan\vartheta_0,-\cot\vartheta_0)$, while keeping $\vartheta_0$ fixed, gives a second linearized solution, whose complex transverse coefficient is $e^{i\sigma(t)}$. These two solutions are independent modulo the flow direction. Hence, every nonzero transverse linearized solution has complex coefficient
\begin{equation}\label{eq: RK-transverse-solution}
 z(t)=e^{i\sigma(t)}\bigl(A+btB+iB\bigr),
 \qquad (A,B)\in\mathbb R^2\setminus\{(0,0)\},
\end{equation}
where $A,B$ are determined by the initial vector. Thus the transverse flow is a uniform rotation of angular speed $2h$, followed by a shear. The shear may reduce the angular speed, and the next estimate shows that its contribution is never large enough to reverse the rotation.

For a continuous argument $\theta(t)=\arg z(t)$, we therefore obtain
\begin{equation}\label{eq: RK-angular-bound}
 \begin{split}
 \dot\theta(t)
 &=2h-\frac{bB^2}{(A+btB)^2+B^2}\ \geq\ 2h-b\\
 &=\frac8{9h}\left(
 \frac{(-2E)^3-27}{(2E+3h^2)^2}+2E+3h^2\right)>0.
 \end{split}
\end{equation}
To obtain the second line, we substitute the relation defining $h$ on the torus. The last identity uses $\cos(2\vartheta_0)=(E+\frac12h^2)h$. Positivity follows from $E<-\frac32$ and $2E+3h^2=\frac{2(h^3+\cos(2\vartheta_0))}{h}>0$, since $h>1$. On the two circular orbits, direct linearization gives angular velocity $2h>0$. The same estimates hold at $E=-\frac32$ away from $\mathcal C$, where $h>1$, and give a uniform positive lower bound on every compact subset of the regular locus.

Finally, let $P=(\xi,T)$ be a noncircular closed orbit on $\mathbb S_2$. Both coordinates are nonzero, so their winding numbers satisfy $(h+h^{-2})T=2\pi k_1, (h-h^{-2})T=2\pi k_2, k_1,k_2\in\mathbb Z_{>0}$.
In \eqref{eq: RK-transverse-solution}, the factor $A+btB+iB$ has a bounded continuous argument. It lies on a horizontal line with a nonzero imaginary part when $B\ne0$, and is a nonzero constant when $B=0$. Thus $\theta(t)=2ht+O(1)$ as $t\to+\infty$, and
\[
 \rho(P)
 =\frac{T}{2\pi}\lim_{t\to+\infty}
   \frac{\theta(t)-\theta(0)}{t}
 =\frac{2hT}{2\pi}
 =k_1+k_2>1.
\]
In particular, every noncircular closed orbit on this double cover has rotation number greater than $1$ and Conley--Zehnder index at least $3$. The explicit positivity estimate will also be used on compact subsets of the critical Kepler level, away from its circle of equilibria.

\subsection{Uniform estimates near the saddle-centers}\label{sec: index of complementary part}
Index growth near a saddle-center equilibrium was established in \cite{GrottaRagazzoSalomao06}. Here we need a version uniform in the small mass and energy parameters. To conclude that a whole periodic orbit has large index, we must also control its complementary segment. This is where the special kinetic form of our regularized Hamiltonians is useful.

We use the standard comparison and concatenation properties of symplectic indices \cite{Long02}, together with the global estimate from \cite[Proposition~7.1]{LS25}. For a Hamiltonian of the form $\mathcal H(y,x)=\frac12|y+F(x)|^2+V(x)$,
the ambient linearized path along any trajectory segment has Robbin-Salamon (RS) index at least $-9$, where the path is started at an arbitrary matrix in $\Sp(4)$. The reason for this bound is simple. Compare the Robbin--Salamon index with the Maslov index of the graph of the linearized path relative to the fixed Lagrangian plane $V\oplus V$, where $V=\R^2\times0$. The condition $\mathcal H_{yy}=I_2$ makes the crossing forms for the latter index positive. The difference between the two indices is an endpoint correction, uniformly bounded in dimension four. This is precisely what gives the constant $-9$.
This lower bound applies both to $\hat H_{\mu,E}$ and to $\hat K_{r,c}$.

We emphasize that the bound is applied to the whole complementary segment, no matter how long this segment is. We start its linearized path at the endpoint matrix of the selected segment and apply the preceding estimate once. Consequently, a large positive contribution from the selected segment cannot be canceled by an arbitrarily long path elsewhere in the chamber. Passing from the ambient path to the transverse path in $\mathrm{Sp}(2)$ changes the index by a fixed bounded amount. The same is true after a local symplectic change of coordinates.

Note that we always compute the index in the global trivialization of the contact structure. In fact, the flow-energy block contributes only a bounded term, while the other block is the transverse linearized flow on the contact structure. Thus, all frame changes and endpoint terms used below are bounded by a constant independent of the length of the chosen segment.

We can now isolate the local estimate that will be used at both necks. The neighborhoods are chosen in the limiting coordinates. The parameter range is reduced only afterward. This order of choices is essential for uniformity.

\begin{lem}\label{lem: uniform-saddle-index}
Given $N>0$, one can choose small neighborhoods $\mathcal U_{i,\pm}$ of $S_{i,\pm}$ and $r_0,\epsilon_0>0$ such that every contractible periodic orbit in the chambers, other than the boundary Lyapunov orbits and their covers, which enters one of the corresponding neighborhoods of $S_{i,\pm}(r,c)$ has $\CZ>N$ for $0<r<r_0$ and $|c-\frac92|<\epsilon_0$. The same statement holds for regular periodic orbits on the critical levels.
\end{lem}
\begin{proof}
The two eigenvalues of positive real and imaginary part at the limiting saddle-center $S_{i,\pm}$ are
\begin{equation}\label{eq: normal-frequencies}
 \lambda_h=\sqrt{\sqrt7+\frac12},\qquad
 \lambda_c=\sqrt{\sqrt7-\frac12}.
\end{equation}
These frequencies remain uniformly positive when the parameters are small. The analytic saddle-center normal form, centered at the moving critical point $S_{i,\pm}(r,c)$, is
\begin{equation}\label{eq: exact-normal-form}
 \hat K_{r,c}=k_i(r,c)+F_{r,c}(I,J),\qquad
 I=\frac{w_1^2-w_3^2}{2},\quad J=\frac{w_2^2+w_4^2}{2},
\end{equation}
where $F=\lambda_h(r,c)I+\lambda_c(r,c)J+O((|I|+J)^2)$.
We use the analytic symplectic version of the normal form theorem \cite{Giorgilli01}, see also \cite{Moser58,dPS1}. The estimates in the convergent normal form construction can be chosen uniformly. In particular, $\partial_IF$ and $\partial_JF$ are positive on a common real neighborhood.

In the variables $u=w_1+w_3$ and $v=w_1-w_3$, Hamilton's equations are $\dot u=\partial_IFu$ and $\dot v=-\partial_IFv$. As $\partial_IF>0$, the variable $u$ increases exponentially and $v$ decreases exponentially along every noncentral trajectory. Therefore, a periodic orbit contained entirely in the normal form neighborhood must lie in $u=v=0$. These are precisely the local Lyapunov orbits. Any other periodic orbit that enters a small ball of radius $d$ must eventually exit a larger ball of fixed radius $D$. The bound $|X_{\hat K}(w)|\leq C|w|$ shows that one such segment has length at least $C^{-1}\log(\frac Dd)$.

This long stay is useful because the center variables rotate with angular speed $\partial_JF$, uniformly bounded below. More precisely, on the larger ball the Hessian in normal form coordinates is uniformly close to $B_0=\operatorname{diag}(\lambda_h,\lambda_c,-\lambda_h,\lambda_c)$ and gives a linear lower bound for the index of a segment of length $T$. Its hyperbolic block has a bounded index, whereas its positive elliptic block has frequency at least a fixed constant $\nu>0$. Thus the segment has index at least $\nu \frac T{\pi}-C_0$, with constant $C_0$ independent of the parameters. The endpoint change from normal form coordinates is bounded because their derivatives lie in a fixed contractible neighborhood of the linear changes.

The complement part of the periodic orbit outside this long segment contributes at worst the fixed negative constant discussed above. Passing from the ambient index to the transverse index also changes the estimate by a bounded amount. Thus the index of the full orbit tends to infinity as $d\downarrow0$, uniformly in the parameters. Choose $\mathcal U_{i,\pm}$ as the small ball of radius $d$ so that this lower bound exceeds $N$, and then reduce the parameter neighborhood of $(r,c)$. This reasoning also applies to regular periodic orbits on the critical level.
\end{proof}

\subsection{The Moon-side index estimate}
\begin{prop}\label{prop: Moon-index}
For some $\epsilon_0,\mu_*>0$, every contractible periodic orbit in the interior of $\mathcal M^m_{\mu,E}$, which is not a Lyapunov orbit, has $\rho>1$ whenever $0<\mu<\mu_*$ and $E<L_1(\mu)+\epsilon_0\eta^2$.
\end{prop}

\begin{proof}
Recall from \eqref{equ: bar E and c} that $E=L_1(\mu)+(\frac92-c)\eta^2$. Thus the desired energy range is equivalent to $c>\frac92-\epsilon_0$. Larger $c$ means lower physical energy. We divide the argument into three overlapping ranges: very low energy, a neighborhood of the critical Hill value, and the compact interval between them. First choose $C>\frac92$ from Lemma~\ref{lem: Kepler-tail}. That lemma proves the assertion for every $c\geq C$, after reducing $\mu_*$ if necessary.

We next treat a full neighborhood of the critical Hill value $c=\frac92$. Let $\mathcal M_{0}^{\rm Hill}$ denote the bounded critical regularized Hill component. Apply Lemma~\ref{lem: uniform-saddle-index} with $N=3$, and then fix the corresponding small disjoint neighborhoods $\mathcal U$ of the four saddle-centers of $\mathcal M_{0}^{\rm Hill}$. After reducing the parameter neighborhood, every contractible periodic orbit that enters $\mathcal U$, other than a boundary Lyapunov orbit, has $\CZ>3$. Hence, a periodic orbit with $\rho\leq1$ must stay in the complement of $\mathcal U$.
The set $\mathcal M'_0=\mathcal M_{0}^{\rm Hill}\setminus\mathcal U$
may be chosen compact and is contained in the regular region of the critical component. By Proposition~\ref{prop: Hill convexity}, $\mathcal M_{0}^{\rm Hill}$ has positive curvature at every point of $\mathcal M'_0$.
Choose a compact neighborhood $\mathcal U'$ of $\mathcal M'_0$ that remains disjoint from the saddle centers. By Proposition~\ref{prop: smooth continuation}, after shrinking $|c-\frac92|$ and $r$, the corresponding portions of the levels of $\hat K_{r,c}$ outside $\mathcal U$ are contained in $\mathcal U'$ and converge to $\mathcal M'_0$ in $C^2$ local graphs as $(r,c)$ tends to $(0,\frac92)$. Hence, their curvatures are also positive, after another reduction of the parameter neighborhood. The usual convexity argument therefore gives, on this fixed compact region, a positive transverse frame whose angular speed has a uniform lower bound $a>0$, independent of the parameters $(r,c)$.

Suppose, by contradiction, that there exist $r_n\to0$, $c_n\to\frac92$ and contractible periodic orbits $P_n$ in the Moon chamber, distinct from the boundary Lyapunov orbits and their covers, such that $\rho(P_n)\leq1$. By the saddle-center estimate above, $P_n$ avoids $\mathcal U$ for $n$ large. The positive transverse frame then gives $T_n\leq \frac{2\pi}a$, where $T_n$ is the period of $P_n$. Since the limiting vector field has no zeros on $\mathcal U'$, there is also a common positive lower bound for $T_n$. Passing to a subsequence, the initial points and periods converge, and the orbit equations and their linearizations converge smoothly. We obtain a nontrivial closed orbit $P_0$ on the critical Hill component, contained in $\mathcal M'_0$. By Proposition~\ref{prop: Hill convexity} and \cite[Corollary~2.7]{LS26}, the limiting orbit $P_0$ has $\rho(P_0)>1$ and $\CZ(P_0)\geq3$. Since the linearized paths of $P_n$ converge to that of $P_0$, $\CZ(P_n)\geq3$ for all sufficiently large $n$. This contradicts $\rho(P_n)\leq1$. Consequently, there are $d_0>0$ and $r_0>0$ such that the desired index estimate holds whenever $0<r<r_0$ and $|c-\frac92|<d_0$.

It remains to cover the energies between the critical neighborhood and the very low-energy range. There is no singular level in this intermediate interval, so ordinary compactness and strict convexity suffice. Choose $0<d<\frac{d_0}2$. For $\frac92+d\leq c\leq C$, the limiting Hill energy levels form a compact family of regular strictly convex hypersurfaces. Since this family remains at a fixed positive distance from the critical Hill level, compactness gives a common positive lower bound for their curvatures. Proposition~\ref{prop: smooth continuation}, used uniformly for $c\in[\frac92+d,C]$, preserves strict convexity for every $r>0$ sufficiently small. The convexity theorem of \cite{HWZ98} then gives $\rho>1$ for every contractible periodic orbit in this range.

Finally, choose $\epsilon_0<\frac{d_0}2$ and reduce $\mu_*$ so that all the preceding estimates hold simultaneously. The critical neighborhood covers $\frac92-\epsilon_0<c<\frac92+d_0$, the uniformly regular convex range covers $\frac92+d\leq c\leq C$, and Lemma~\ref{lem: Kepler-tail} covers $c\geq C$. Since these intervals overlap, they cover every $c>\frac92-\epsilon_0$, which is exactly the range $E<L_1(\mu)+\epsilon_0\eta^2$. This proves the proposition.
\end{proof}

\subsection{The Earth-side index estimate}\label{sec: Earth-index}
The Earth chamber requires an additional argument. The limiting rotating Kepler system has a whole circle of equilibria $\mathcal C$ at energy $-\frac32$. We distinguish two possibilities for a periodic orbit near this circle. Either the orbit enters the first neck, where the saddle-center index estimate in Lemma \ref{lem: uniform-saddle-index}
applies, or it stays away from the neck and makes a slow passage through a fixed angular sector. In the second case, the Kepler frame gives a large index.

The invariant torus $T_{\mu,E}$ separates the Earth chamber into a compact region $\mathcal T^c$ containing the retrograde orbit and a region $\mathcal T$ near $\mathcal C$. Below $L_1$, the latter is a solid torus. Above $L_1$, it is a solid torus with the neck ball removed. The invariant torus from Section~\ref{subsec: Rotating Kepler problem} separates the region where the Kepler compactness argument works from a thin tube near $\mathcal C$. Inside this tube we show that a periodic orbit must either visit the neck or travel across a fixed angular sector opposite the Moon. Such a passage is very slow, and the linearized elliptic direction produces a large index. As $(\mu,E)\rightarrow (0,-\frac32)$, the torus $T_{\mu,E}$ converges uniformly to $T_0$ in the rotating Kepler problem.

Write $q=\rho e^{i\theta}$ in the configuration plane, with $\theta=\pi$ corresponding to the point $q=-1$ on $\mathcal C$. By an angular sector about $q=-1$, we mean a set of the form $\Sigma_-=\{(\rho,\theta): |\theta-\pi|\leq \theta_1\}$ intersected with the fixed tubular neighborhood of $\mathcal C$ under consideration. The two angular sides of the chosen sector must remain a fixed positive distance apart in configuration space. We therefore choose a smaller sector $\Sigma_-^0=\{(\rho,\theta): |\theta-\pi|\leq \theta_0\}, 0<\theta_0<\theta_1$, with closure contained in the interior of $\Sigma_-$. We say that an orbit
makes a full passage through $\Sigma_-$ if one connected component of
its intersection with $\Sigma_-$ enters through one angular side of $\Sigma_-^0$ and leaves through the other.

\begin{lem}\label{lem: slow-circle-index}
Fix an angular sector $\Sigma_-$ about $q=-1$ and a smaller sector $\Sigma_-^0\subset\operatorname{int}\Sigma_-$. Given $N>0$, choose the two-torus $T_0$
sufficiently close to $\mathcal C$ and then the parameter neighborhood
sufficiently small. Then every contractible periodic orbit in $\mathcal T$ which makes a full passage through $\Sigma_-$ satisfies $\CZ>N$.
\end{lem}

\begin{proof}
At $(p,q)=(-ie^{is},e^{is})\in\mathcal C$, the Hessian of $H_0$ is
\[
 \begin{pmatrix} I_2&J_2\\J_2^T&M_s\end{pmatrix},\qquad
 M_s=\begin{pmatrix}1-3\cos^2s&-3\cos s\sin s\\
 -3\cos s\sin s&1-3\sin^2s\end{pmatrix},
 \quad J_2=\begin{pmatrix}0&-1\\1&0\end{pmatrix}.
\]
The following fixed linear transformation
\[
 V_s=\begin{pmatrix}
 -\sin s&\sin s&\cos s&\cos s\\
 \cos s&-\cos s&\sin s&\sin s\\
 \cos s&-2\cos s&-2\sin s&-\sin s\\
 \sin s&-2\sin s&2\cos s&\cos s
 \end{pmatrix}
\]
puts the Hessian of $H_0$ in the diagonal form $\operatorname{diag}(1,-3,1,0)$. Use the fixed frame $V_\pi$ on a sufficiently narrow sector around $q=-1$. There, the Hessians of $H_\mu$ are uniformly close to this matrix when $\mathcal T$ is close to $\mathcal C$ and $\mu$ is small. For some $\tau>0$ small, comparison with $\operatorname{diag}(1-\tau,-3-\tau,1-\tau,-\tau)$ gives a positive index growth. Its first symplectic block is positive elliptic with frequency $1-\tau$, while the second is negative elliptic with frequency $\sqrt{\tau(3+\tau)}$. Choose $\tau$ small enough so that $1-\tau-\sqrt{\tau(3+\tau)}>0$. Then the positive elliptic block rotates much faster than the negative block. Our choice of $\tau$ leaves a strictly positive net rotation. The comparison theorem then gives an ambient index lower bound $bT-C$ for every segment of length $T$ contained in the fixed sector, with $b>0$ independent of the small parameters $r,E+\frac32$.

Shrinking the tube $\mathcal T$ around $\mathcal C$, the velocity $\dot q=p+iq$ tends uniformly to zero on $\Sigma_-$. The two sides of the smaller sector $\Sigma_-^0$ are separated near $\mathcal C$ by a  fixed small distance $d_0>0$. Hence, a segment which makes a full passage through $\Sigma_-$ has length in time at least $T\geq \frac{d_0}{\delta}$, where $\delta$ is the supremum $|\dot q|$ on $\mathcal T\cap\Sigma_-$. This lower bound tends to $+\infty$ as the tube and the parameter neighborhood shrink. The preceding linear lower bound of index for the selected segment, the complementary-segment estimate of Section \ref{sec: index of complementary part} give $\CZ>N$. The coordinate changes to the regularized frame make only bounded corrections to their index, which is independent of the time spent in the sector.
\end{proof}

The preceding lemma gives a large index provided the orbit crosses the slow sector. We still need to show that an orbit cannot remain forever in the small angular wedge facing the Moon while avoiding the neck. This is the role of the next lemma.

\begin{lem}\label{lem: Earth-alternative}
Fix the inner neck neighborhoods in Lemma~\ref{lem: uniform-saddle-index}. For $\epsilon_0,\mu_*>0$ small enough, every periodic orbit contained in $\mathcal T$, other than the boundary Lyapunov orbit and its covers, either enters the inner neck neighborhoods over $l_1$ or makes a full passage through the sector in Lemma~\ref{lem: slow-circle-index}. This holds for $E\leq L_1(\mu)+\epsilon_0\eta^2$ sufficiently close to $-\frac32$, including critical regular orbits.
\end{lem}

\begin{proof}
Write $q+\mu=\rho e^{i\theta}, s=1-\rho$, and $\kappa=-E-\frac32$.
Fix $c_0\in\left(\frac32,\frac92\right)$. Since $\frac{\kappa}{\eta^2}\to\frac92$ at the first critical value, we may reduce
the rescaled energy window and then take $\mu$ sufficiently small so that
\begin{equation}\label{eq: kappa-lower}
 \kappa\geq c_0\eta^2,\qquad
 \mu=O(\kappa^{3/2}).
\end{equation}

In polar coordinates $(\rho,\theta)$, the effective potential is
\[
U_\mu=-\frac{1-\mu}{\rho}
-\frac{\mu}{\sqrt{1+\rho^2 -2\rho\cos\theta}}-\frac12\big(\rho^2-2\mu\rho\cos\theta+\mu^2\big).
\]
Let $d=|\rho e^{i\theta}-1|$ be the distance from the particle to the Moon in Earth-centered
configuration coordinates. Since $H_\mu=\frac12|v|^2+U_\mu$ with $v=\dot q$, the energy equation gives
\begin{equation}\label{eq: Earth-speed-exact}
|v|^2=s^2\frac{2+\rho}{\rho}-2\kappa
+2\mu\left(\frac1d-\frac1\rho-\rho\cos\theta\right)+\mu^2.
\end{equation}
On every fixed angular sector away from $\theta=0$, we have $d\geq d_*>0$, and hence
\begin{equation}\label{eq: Earth-speed}
|v|^2=s^2\frac{2+\rho}{\rho}-2\kappa
+O(\mu),
\end{equation}
with a uniform remainder. The Earth branch has $s>0$. On a sufficiently
small tube, \eqref{eq: kappa-lower} and \eqref{eq: Earth-speed} imply $|v|\leq s\sqrt{\rho^{-1}(2+\rho)}$
on every such fixed sector after reducing $\mu_*$.

Recall $h^2=-2E+2\operatorname{Im}(\bar qp)$ from Section \ref{subsec: Rotating Kepler problem}. Define the circle-valued function coming from McGehee's coordinates by
\begin{equation}\label{eq: theta2-near-C}\begin{split}
\theta_2=\arg\left( q^{1/2} \left(\bar p+ih\frac{\bar q}{|q|}\right) \right) -\frac{q\bar p+\bar qp}{4h^2}\quad\in\R/\pi\Z.
\end{split}
\end{equation}
It is smooth as an $\R/\pi\Z$-valued function on the tube for each $\mu>0$. Changing the branch of $q^{1/2}$ changes the argument by $\pi$, thus the class modulo $\pi$ is well defined. Moreover, $h$ is close
to $1$, and the other complex factor is nonzero. Uniformly on the tube,
\begin{equation}\label{eq: theta2-angle-close}
\theta_2=\frac\pi2-\frac\theta2+o(1)
\qquad\text{in }\R/\pi\Z.
\end{equation}
On every closed angular sector $\Sigma_-'$ disjoint from the Moon direction,
its derivative satisfies
\begin{equation}\label{eq: theta2-positive}
2|q|X_{H_\mu}\theta_2= h-\frac1{h^2}+O(\mu).
\end{equation}
Indeed, this is the exact rotating Kepler formula at $\mu=0$, and the
claim follows from the $C^1$-closeness of the vector fields and of \eqref{eq: theta2-near-C} on $\Sigma_-'$. The factor $2|q|$ is
the Levi-Civita time factor for the convention used in
\eqref{equ: motion on the round 3-sphere}.

Since $p=v-iq$, from the definition of $h$, we obtain $h^2 \geq 3+2\kappa-2\rho^2 -2\rho|v|-C\mu$. Using the previous upper bound for $|v|$ and expanding at $\rho=1$, we obtain,
after reducing the tube, $h^2 \geq 1+2\kappa+(2-\sqrt3)s-C\mu$.
Together with \eqref{eq: kappa-lower} and
\eqref{eq: theta2-positive}, this implies $X_{H_\mu}\theta_2>0$
on every fixed closed angular sector away from the Moon.

We first use $\theta_2$ to decide whether the orbit winds around the critical circle. The key point is that on every angular sector separated from the Moon direction, $\theta_2$ increases strictly along the flow. A periodic orbit that reaches one of these sectors must therefore traverse all angle values. If it does not, it remains trapped in a narrow wedge directed toward the Moon. We shall rule out this last possibility outside the neck.
Since $2c_0>3$, choose $c_1>1$ sufficiently close to $1$ so that $3c_1^2<2c_0$. Then choose $0<\theta_*'<\theta_*$
small enough so that $\cos\theta_*>c_1^{-3}$. Reduce the tube and the parameter neighborhood so that the following consequence of \eqref{eq: theta2-angle-close} holds: if a value $\vartheta\in\R/\pi\Z$ has one preimage in the tube with $|\theta|\geq\theta_*$, then every preimage of $\vartheta$ in the tube lies in a small angular sector centered at the same angle and disjoint from $|\theta|\leq\theta_*'$. On this entire preimage set, $X_{H_\mu}\theta_2>0$.

Let $P$ be a periodic orbit in the tube. If $P$ is not contained in the
wedge $|\theta|<\theta_*$, choose a value $\vartheta\in\R/\pi\Z$ attained by $P$ at a point with $|\theta|\geq\theta_*$. Every crossing at $\theta_2=\vartheta$ is positive. Therefore $\theta_2|_P: P\simeq S^1\longrightarrow\R/\pi\Z$ has positive degree and is onto.

Now use \eqref{eq: theta2-angle-close} near the sector $\Sigma_-$ about $q=-1$. Choose a closed interval $I_-=[\vartheta_-,\vartheta_+] \subset\R/\pi\Z$ whose preimage near $q=-1$ lies in $\Sigma_-$ and contains $\Sigma_-^0$. Choose its endpoints so that the levels $\vartheta_-$ and $\vartheta_+$ lie on opposite angular sides of $\Sigma_-^0$. Since $X_{H_\mu}\theta_2>0$ there, every component over $I_-$ is crossed in the positive direction. Because the degree is positive, at least one component of $P$ over $I_-$ goes from $\vartheta_-$ to $\vartheta_+$. Hence it makes a full passage
through $\Sigma_-$. Thus every periodic orbit in the tube either makes this slow passage or stays inside the wedge $|\theta|<\theta_*$.

It remains to exclude a periodic orbit contained entirely in this Moon-facing wedge and avoiding the neck. We shall find a function which is strictly monotone along such an orbit. The required estimate is $\partial_{q_1}U_\mu>0$
throughout this wedge, see also \cite{Birkhoff1915}. Suppose first that $s\geq c_1\eta$.
Direct differentiation gives
\begin{equation}\label{eq: Earth-wedge-derivative}
 \partial_{q_1}U_\mu
 \geq
 (3-o(1))s\cos\theta_*
 -\frac{\mu}{s^2}
 -O(\mu).
\end{equation}
Here we used $|q-(1-\mu)|\geq s$ and $\mu s^{-2} \leq 3c_1^{-2}\eta$, since $\mu=3\eta^3$. Therefore, the leading coefficient in
\eqref{eq: Earth-wedge-derivative} is positive by the choice $\cos\theta_*>c_1^{-3}$.
Thus $\partial_{q_1}U_\mu>0$
for all sufficiently small parameters.

It remains to consider the region $s<c_1\eta$.
We show that, on the allowed energy window, this strip lies arbitrarily close to the first Hill saddle. This is why the restriction to a small energy window is necessary. After choosing $c_1>1$ sufficiently close to $1$ and reducing the rescaled energy window, every allowed Earth-side point in this region lies in the prescribed inner neck neighborhood in Lemma \ref{lem: uniform-saddle-index}.

Suppose otherwise, and take a sequence with $\mu\to0$ remaining outside
that neighborhood. If $\frac{|\theta|}{\eta}\to\infty$,
then $\frac d\eta\to\infty$.
Dividing \eqref{eq: Earth-speed-exact} by $\eta^2$ and using $\mu=3\eta^3$, we obtain $0 \leq \eta^{-2}|v|^2 \leq 3c_1^2-2c_0 +o(1)$, which contradicts $3c_1^2 <2c_0$. Hence $\theta=O(\eta)$ in the energy surface.

Introduce Moon-centered Hill coordinates $Q=Q_1+iQ_2=\eta^{-1}(\rho e^{i\theta}-1)$.
Since $\theta=O(\eta)$, $Q_1 = -\frac{s}{\eta}+O(\eta)$ and $Q_2 = \frac{1}{\eta}\rho\sin\theta = O(1)$. Thus the Hill-scaled configurations remain in a bounded set. Write $Q_1=-\ell+o(1)$.
The condition $s<c_1\eta$ gives $\ell\leq c_1+o(1)$.
The side selection can be justified using the intermediate section $s=\eta$. On this section, the same energy estimate gives $\theta=O(\eta)$ and hence $Q_1=-1+O(\eta)$. Moreover, \eqref{eq: Earth-speed-exact} and the assumed energy bound give $0\leq |\frac v\eta|^2\leq -6+2\epsilon_0+\frac6{|Q|}+o(1)$. Since $|Q|\geq1-o(1)$, we obtain $Q\to(-1,0)$ and $\frac v\eta\to0$ as $\epsilon_0,\mu\to0$. Thus, after reducing $\epsilon_0$ and $\mu_*$, the entire allowed part of this section lies in the prescribed inner neck neighborhood. The saddle-center normal form shows that removing this normal-form ball leaves the Earth-side piece connected. This piece contains the Earth core and does not meet $s=\eta$, so it lies on the side $s>\eta$. Consequently, $\ell\geq1-o(1)$. All subsequent reductions of $\epsilon_0$ and $\mu_*$ preserve this separation.
Therefore
\begin{equation}\label{eq: ell-near-one}
 1-o(1)\leq\ell\leq c_1+o(1).
\end{equation}

Let $c_H=-\bar E$ denote the Hill energy parameter and put $R=|Q|$.
Since $R\geq |Q_1|=\ell+o(1)$,
the Hill-scaled energy equation takes the form $|V|^2 = -2c_H+3\ell^2+\frac6R+o(1)$.
By nonnegativity of the left-hand side, $c_H \leq \frac12 ( 3\ell^2+\frac6\ell) +o(1)$. The function $f(\ell) := \frac12 ( 3\ell^2+\frac6\ell)$ has a strict minimum $f(1)=\frac92$ at $\ell=1$. Hence \eqref{eq: ell-near-one} gives $c_H \leq \frac92+o(|c_1-1|)+o(1)$. On the other hand, the assumed rescaled energy window $E\leq L_1(\mu)+\epsilon_0\eta^2$
implies $c_H \geq \frac92-\epsilon_0+o(1)$. Thus, by first choosing $c_1>1$ sufficiently close to $1$ and then $\epsilon_0>0$ sufficiently small, every such point lies on a Hill
energy level arbitrarily close to the first critical level.

At the critical Hill energy $c_H=\frac92$, the energy equation becomes $|V|^2 = -9+3\ell^2+\frac6R$. For $1\leq\ell\leq c_1$, and $R\geq\ell$, nonnegativity gives $\ell \leq R \leq \frac6{9-3\ell^2}$. Consequently, $Q\to(-1,0)$ and $V\to 0$ uniformly as $c_1\downarrow1$. Choose $c_1$ sufficiently close to $1$ so that this portion of the critical Hill level is contained in the prescribed inner neck neighborhood. By the preceding control of $c_H$ and smooth dependence of the Hill-scaled levels, the same inclusion
holds throughout every sufficiently small rescaled energy window. Finally, by the smooth convergence of the rescaled restricted problem to Hill's problem established in Section \ref{sec: smooth continuation}, it holds for all sufficiently small $\mu$.

We have proved that the only points of the allowed energy level in the thin strip belong to the chosen inner neck neighborhood. Consequently, every periodic orbit confined to the wedge and avoiding the neck satisfies $\partial_{q_1}U_\mu>0$ throughout its trajectory. Since $\dot p_1+\dot q_2 = -\partial_{q_1}U_\mu$, the function $p_1+q_2$ would be strictly decreasing along the whole periodic orbit, which is impossible. Hence every periodic orbit in the tube, other than the boundary Lyapunov orbit and its covers, either enters the prescribed inner neck neighborhood or makes a full passage through $\Sigma_-$. This proves the lemma.
\end{proof}
The two alternatives are complete. An orbit in the outer tube either approaches the saddle-center or spends a long time in the fixed slow sector. In both cases its index is large. What remains is to combine this conclusion with the compact rotating Kepler region inside the invariant torus. Notice that
excluding periodic orbits of rotation number $1$ is enough for the foliation criterion in Theorem \ref{thm: criterion of FEF}.

We now complete the Earth-side estimate.
\begin{prop}\label{prop: Earth-index}
There exist $\epsilon_0,\mu_*>0$ such that every contractible periodic orbit in the interior of the Earth chamber, other than a boundary Lyapunov orbit, has $\rho>1$ whenever $0<\mu<\mu_*$ and $E<L_1(\mu)+\epsilon_0\eta^2$.
\end{prop}

\begin{proof}
We first consider energies sufficiently far below $-\frac32$. For $E\leq-K$, the conclusion follows from Lemma~\ref{lem: Kepler-tail}. On every compact interval of energies below $-\frac32$, it follows from the smooth convergence to the bounded rotating Kepler component and its dynamical convexity explained in Section \ref{sec_positive_frame}.

We now consider energies converging to $-\frac32$. Fix the invariant torus $T_{\mu,E}$ constructed in
Section~\ref{subsec: Rotating Kepler problem}, and write the Earth chamber as the union of the compact part $\mathcal T^c$ bounded by this torus and the remaining tube $\mathcal T$ near the critical circle $\mathcal C$. We choose the limiting torus $T_0$ sufficiently close to $\mathcal C$ to apply Lemma~\ref{lem: slow-circle-index} with an index threshold larger than $3$.

On $\mathcal T^c$ the limiting rotating Kepler flow is regular, and the positive frame of Section~\ref{sec_positive_frame} has uniformly positive angular speed. The same holds for the nearby flows. The following argument is similar as in the proof of Proposition \ref{prop: Moon-index}. Suppose that a sequence of contractible periodic orbits in $\mathcal T^c$ had rotation number at most $1$. The angular-speed bound would bound their periods from above. Since the vector fields have no zeros on this compact region, the periods are also bounded away from zero. Passing to a subsequence gives a nonconstant closed rotating Kepler orbit. The explicit calculation in Section \ref{sec_positive_frame}, including its two circular orbits, gives a rotation number $\rho>1$ for the limiting orbit. The lower semicontinuity of the Conley-Zehnder index contradicts the assumed bound on the nearby orbits.

Now let $P$ be a periodic orbit in tube $\mathcal T$. By Lemma~\ref{lem: Earth-alternative}, either $P$ enters a prescribed inner neck neighborhood or it makes a full passage through the slow sector. Lemmas~\ref{lem: uniform-saddle-index} and \ref{lem: slow-circle-index} give $\CZ(P)>3$, respectively. In particular $\rho(P)>1$. Finally, the invariant torus itself has irrational rotation number, so
it contains no periodic orbit. The two cases therefore exhaust all
periodic orbits in the Earth chamber near the critical energy. Combining this with the lower-energy argument and reducing $\epsilon_0,\mu_*$ once
more proves the proposition.
\end{proof}

\section{Contact forms and dividing spheres}\label{sec: contact}
We want to apply the finite energy foliation criterion Theorem \ref{thm: criterion of FEF} to the Earth and Moon chambers. Hence, we must first put their Hamiltonian flows into the contact setting required by that criterion.  To do that, we construct the contact form from a Liouville vector field transverse to the regularized energy level. Away from the necks there are natural choices, one on the Earth side and another on the Moon side. However, they do not match at the first saddle-center. In fact, after the Moon rescaling, the Earth field has size of order $r^{-1}$ while the Moon field remains bounded. In this section, we join them without losing transversality. Once this is done, the dividing holomorphic hemispheres are obtained almost explicitly from the local saddle-center normal form.

We consider energies in the range $E=L_1+\epsilon\eta^2$, $0<\epsilon<\epsilon_0$, uniformly for small $\mu$. The constructions below work on both sides of $L_2$, whenever the corresponding dividing sphere exists. The constants in the local estimates do not depend on $\mu$, after $\mu$ and $\epsilon_0$ have been made sufficiently small.

\subsection{An interpolation lemma}

Fix one of the saddle-centers and the analytic symplectic coordinates given by the normal form in Section \ref{sec: index-estimates}. The Hamiltonian is $k_i+F(I,J)$, where $I=\tfrac12(w_1^2-w_3^2), J=\tfrac12(w_2^2+w_4^2)$, and $\partial_IF,\partial_JF$ are strictly positive on a common neighborhood. The energy level is $F(I,J)=\delta$, with $\delta=\h(c)-k_i(r,c)$.

In these coordinates, consider $\omega_0=dw_1\wedge dw_3+dw_2\wedge dw_4$ and the Liouville vector field
\begin{equation}\label{eq: local-Y1}
 Y_1=\frac12(3w_1\partial_{w_1}+w_2\partial_{w_2}
             -w_3\partial_{w_3}+w_4\partial_{w_4}).
\end{equation}
It satisfies
\begin{equation}\label{eq: Y1-positive}
 d\hat K(Y_1)=\frac{\partial_IF}{2}(3w_1^2+w_3^2)
+\frac{\partial_JF}{2}(w_2^2+w_4^2)\geq b_0|w|^2
\end{equation}
for a uniform $b_0>0$. Thus $Y_1$ is transverse to every nearby regular level, on either side of the critical value $L_2$.

Since $\partial_IF,\partial_JF$ are positive, the energy equation gives
\begin{equation}\label{eq: correct-cone}
 w_1^2+w_2^2+w_4^2\leq C(\delta^++w_3^2),\qquad
 \delta^+=\max(\delta,0).
\end{equation}
Indeed, the leading part of the equation is $2\delta=\lambda_h(w_1^2-w_3^2) +\lambda_c(w_2^2+w_4^2)+O((|I|+J)^2)$.
On a sufficiently small fixed ball, the error can be absorbed into the positive terms on the left.  In particular, if $|w_3|$ is bounded below by $a>0$ and $\delta^+a^{-2}$ is small, the energy surface lies close to the hyperbolic cone. More precisely, on this region we obtain
\begin{equation}\label{eq: sharper-cone}
 |w_2|\leq\left(\sqrt{\frac{\lambda_h}{\lambda_c}}+o(1)\right)|w_3|.
\end{equation}
The term $o(1)$ can be made small by first shrinking the normal form ball and then reducing $\delta^+a^{-2}$. We use this estimate only in regions bounded away from $w_3=0$.

The next lemma states precisely what we need. We write $s=|w_3|>0$ for the coordinate pointing into one side of the saddle. The large factor $t^{-1}$ is included in the assumptions, because $t=r$ associates to the Earth side, whereas $t=1$ associates to the Moon side.

\begin{lem}\label{lem: two-stage-gluing}
Let $Z$ be a Liouville vector field on one side $s=\sigma w_3>0$ of a saddle-center, where $\sigma\in\{-1,1\}$. Suppose
\begin{equation}\label{eq: primitive-assumption}
 \iota_{Z-Y_1}\omega_0=dG,\quad
 G=\frac1t\big(\sigma A w_1+B w_4+R(w)\big),
 \quad R=O(|w|^2),\quad dR=O(|w|),
\end{equation}
where $0<t\leq1$, $A>0$, all coefficients and remainder bounds are uniform, and
\begin{equation}\label{eq: domination}
 A\sqrt{\lambda_h}>|B|\sqrt{\lambda_c}.
\end{equation}
Then there are $0<a<b$ and $\delta_0>0$ such that, for $|\delta|<\delta_0$, the two fields can be joined by a Liouville field transverse to $F(I,J)=\delta$. It equals $Y_1$ for $0<s\ll a$ and equals $Z$ for $s\geq b$. The constants are independent of $t$.
\end{lem}
\begin{proof}
We use the convention $\iota_{X_G}\omega_0=-dG$, which means $Z=Y_1-X_G$. Choose $b>0$ small and then $a>0$ much smaller. Separate the leading hyperbolic part $G_h=\frac{\sigma A}{t}w_1$
from the rest of $G$. First take a nondecreasing cutoff $\gamma(s)$, zero for $s\leq \frac a4$ and one for $s\geq \frac a2$, and put $Y^{(1)}=Y_1-X_{\gamma G_h}$. Direct computation gives
\begin{equation}\label{eq: first-stage-good-sign}
 d\hat K(Y^{(1)})=d\hat K(Y_1)
 +\frac A t\partial_IF \big(\gamma s+\gamma' w_1^2\big)>0
\end{equation}
at every regular point. Near the saddle-center, we keep $Y_1$, and beyond $s=\frac a2$ we have the correction $-X_{G_h}$.

Choose another nondecreasing cutoff $\chi(s)$ which is zero for $s\leq a$ and one for $s\geq b$, with $0\leq s\chi'(s)\leq\frac{C_\chi}{\log(b/a)}$.
More precisely, we choose $\chi(s)=\chi_0(\frac{\log s-\log a}{\log(b/a)})$ using a fixed standard nondecreasing cutoff $\chi_0$.
By making $\frac ab$ small, the derivative term can therefore be made as small as we wish.
In this second interval of $s$, put
$$Y=Y_1-X_{G_h+\chi(G-G_h)}=(1-\chi)(Y_1-X_{G_h})+\chi Z-(G-G_h)X_\chi.$$
By \eqref{eq: sharper-cone} and \eqref{eq: domination}, both $Z$ and $Y_1-X_{G_h}$ satisfy $d\hat K(Z),\ d\hat K(Y_1-X_{G_h}) \geq b_0|w|^2+\frac{m}{t}s$ for some $m>0$, after reducing $b$ and then $\delta^+a^{-2}$. Here is the reason for the strict inequality in the hypothesis. The leading linear contributions of $d\hat K(-X_G)$ and $d\hat K(-X_{G_h})$ to transversality are $\frac1t(A\lambda_h s+B\lambda_c w_2)$ and $\frac{1}{t}A\lambda_hs$, respectively.
By the cone estimate \eqref{eq: sharper-cone}, the first is at least $\frac{s}{t}\sqrt{\lambda_h} \left(A\sqrt{\lambda_h}-|B|\sqrt{\lambda_c}-o(1)\right)$, which is positive because of \eqref{eq: domination}. The quadratic remainder of $G$ contributes $O(\frac{s^2}t)$, which is smaller after $b$ is chosen sufficiently small. Indeed, there is a uniform margin $\frac{ms}t$ in both endpoint fields. Finally, the extra term in $Y$ is $(G-G_h)X_\chi$. Its absolute contribution $|(G-G_h)d\hat K(X_\chi)|$ is at most $\frac C t s^2\chi'(s) \leq\frac{CC_\chi}{\log(b/a)}\frac s t$. Choose $\frac ab$ small enough that this error is bounded by $\frac {ms}{2t}$. The positive margin just obtained then absorbs it. Only after $a$ and $b$ have been fixed, we choose $\delta_0$ so small that $\delta^+a^{-2}$ lies in the required range for $|\delta|<\delta_0$.

The resulting field agrees with $Y_1$ near the saddle, agrees with the intermediate field in the overlap, and equals $Z$ for $s\geq b$. It is transverse throughout the regular energy level. Since both modifications were made by Hamiltonian fields, the new field is Liouville. All bounds on the large terms contain the same factor $t^{-1}$, so the argument is uniform for $0<t\leq1$.
\end{proof}

\subsection{The Liouville fields on both sides}\label{sec: two-sided estimates}
We now identify the fields that have to be joined. The Earth field is radial in the configuration variable centered at the Earth. The Moon field is the radial field in the rescaled regularized coordinates. Each has a useful transversality property away from the necks, while its difference from $Y_1$ has a primitive of the form required in the previous lemma.

On the Earth side take $Y_e=(q_1+\mu)\partial_{q_1}+q_2\partial_{q_2}$.
The field is Liouville for $\omega_0=\sum dp_j\wedge dq_j$. In the Levi-Civita chart at the Earth collision, see Section \ref{sec: the lower-energy range}, it becomes the radial field $\tfrac12(u\partial_u+v\partial_v)$, and hence extends smoothly through the collision. Hence, our contact form $\iota_{Y_e}\omega_0$ is defined on the whole regularized chamber.

In the rescaled Moon coordinates take $Y_0=\frac12(y_1\partial_{y_1}+y_2\partial_{y_2} +x_1\partial_{x_1}+x_2\partial_{x_2})$.
The Moon field is Liouville and is adapted to the strictly convex Hill energy levels. We first confirm the transversality of the two fields away from fixed small neighborhoods of the saddle-centers. The interpolation in Lemma \ref{lem: two-stage-gluing} will take care of those neighborhoods. The transversality on Earth side requires a separate argument because, at $\mu=0$, the bounded Kepler level develops a whole circle of critical points. We give the details in the next lemma.

\begin{lem}\label{lem: outer-Earth-transverse}
For every fixed choice of sufficiently small Hill-scaled neighborhoods of the first neck, there is $\epsilon_0>0$, independent of $\mu$, such that outside these neighborhoods, $dH_\mu(Y_e)>0$ on the Earth part for all sufficiently small $\mu$ and $E=L_1(\mu)+\epsilon\eta^2, |\epsilon|<\epsilon_0$.
\end{lem}
\begin{proof}
There are three kinds of points to consider: points that remain away from the critical circle $\mathcal C$, points approach $\mathcal C$ but escape a bounded Hill-scaled region, and points that remain in a bounded Hill-scaled region. We prove positivity in each case. With $v=p+iq$ and $q+\mu=\rho e^{i\theta}$, we have $dH_\mu(Y_e)\geq\rho\big(\partial_\rho U_\mu-|v|\big)$.
On compact sets away from $\mathcal C$, positivity follows from the limiting rotating Kepler problem and compactness. At $\mu=0$ and $E\leq-\frac32$ the energy equation gives $|v|^2\leq-3+\frac 2\rho+\rho^2$ on the bounded component, while $\partial_\rho U_0=\frac1{\rho^2}-\rho$. Hence, for $0<\rho<1$, $(\partial_\rho U_0)^2-|v|^2 \geq \frac1{\rho^4}(1-\rho)^2 (3\rho^2+2\rho+1)>0$. Both $\partial_\rho U_0$ and the expression on the right are positive for every $0<\rho<1$. Thus $\partial_\rho U_0>|v|$, and the desired transversality follows at the rotating Kepler limit. On any compact set disjoint from the critical circle the inequality is strict by a uniform amount, so it persists for small $\mu$ and nearby energies. This argument also extends across the regularized Earth collision, where $Y_e$ is radial, see \cite{AFKP12}.

The only possible loss of positivity must therefore occur along a sequence approaching the critical circle $\mathcal C$. Consider a sequence approaching $\mathcal C$ and write $s=1-\rho$. If $\frac s\eta \to\infty$, then $d=|\rho e^{i\theta}-1|\geq s$ and direct differentiation, together with \eqref{eq: Earth-speed-exact}, gives
$\partial_\rho U_\mu\geq(3-o(1))s-\frac\mu{s^2}$ and $|v|^2\leq(3+o(1))s^2+\frac{2\mu}{s}.$
Since $\frac\mu{s^3}\to0$, we have $\partial_\rho U_\mu=(3-o(1))s$ and $|v|\leq(\sqrt3+o(1))s$. Hence $\partial_\rho U_\mu-|v|>0$.

Suppose next that $\frac s\eta$ is bounded but $\frac d\eta \to\infty$. In this case the Moon terms satisfy $\frac{\mu}d=o(\eta^2)$ and $\frac{\mu}{d^2}=o(\eta)$. From the identity \eqref{eq: Earth-speed-exact}, $|v|^2=3s^2-2\kappa+o(\eta^2)$.
Because the left-hand side is nonnegative and, in this two-sided near-critical window, $\kappa\geq c\eta^2$ for some fixed $c>0$, the ratio $\frac s\eta$ is bounded away from zero. Also $\partial_\rho U_\mu=3s+o(s)$ and
$|v|$ is bounded by $(\sqrt3+o(1))s$. The inequality is therefore strict. These two cases exhaust the sequences that leave every compact set in the Hill coordinates.

It remains to consider bounded Hill-scaled configurations near the first saddle. In this region, the separate Earth and Moon terms can no longer be estimated as small perturbations of the rotating Kepler problem. We therefore turn to the limiting Hill problem and first prove the inequality there, excluding a prescribed neighborhood of the equilibrium. Smooth convergence will then transfer it back to the restricted problem.

Start at the critical Hill energy. In Hill coordinates $(P,Q)$, the Earth's Liouville field is
$
Y_e=(\eta^{-1}+Q_1)\partial_{Q_1}+Q_2\partial_{Q_2}.
$
Since $H_\mu=-\frac12(1-\mu)(3-\mu)+\eta^2H_\mu^{\mathrm{Hill}}$, see Section~\ref{sec: Hill's problem}, its derivative along $Y_e$ satisfies
$dH_\mu(Y_e)=\eta\bigl(\partial_{Q_1}U^{\mathrm{Hill}} +V_2+o(1)\bigr).$
Here, we use that $V=V_1+iV_2=P+iQ$ and $U^{\mathrm{Hill}}(Q)=-\frac{3}{|Q|}-\frac32Q_1^2$. Thus, on compact subsets away from the saddle, transversality follows if $\partial_{Q_1}U^{\mathrm{Hill}}>|V|$. Write $Q_1=-\ell\leq-1$ and let $R=|Q|\geq\ell$. Then
\[
\partial_{Q_1}U^{\mathrm{Hill}}=3\ell(1-\frac1{R^3}) \geq3\ell-\frac3{\ell^2},\qquad
|V|^2=-9+3\ell^2+\frac6R \leq-9+3\ell^2+\frac6\ell.
\]
For $\ell>1$ the required strict inequality follows from
\begin{equation}\label{eq: exterior-Hill-polynomial}
(3\ell-\frac3{\ell^2})^2-(-9+3\ell^2+\frac6\ell)
=\frac{3}{\ell^4}(\ell-1)^2(2\ell^4+4\ell^3+9\ell^2+6\ell+3)>0.
\end{equation}
For $\ell=1$, the energy equation forces $R=1$ and $V=0$, so the only point is the equilibrium. Therefore, on every compact part of the critical Earth-facing component outside a prescribed neighborhood of this equilibrium, there is a uniform positive margin in $\partial_{Q_1}U^{\mathrm{Hill}}-|V|$.

Now choose the two-sided rescaled energy window $|\epsilon|<\epsilon_0$ sufficiently small. Smooth dependence of the Hill-scaled energy surfaces preserves this positive margin for all limiting Hill energies in that range. The smooth Hill continuation result in Section \ref{sec: smooth continuation} then gives the same inequality for the rescaled restricted problem when $\mu$ is sufficiently small. Combining the bounded Hill region case with the two escaping cases proves the assertion away from the chosen neck neighborhood.
This completes the proof.
\end{proof}

The corresponding statement on the Moon side is simpler. Recall that the bounded critical Hill component is convex, and the origin lies in the interior of the domain it bounds. At every regular point away from the saddle-centers, the radial field is therefore strictly transverse, i.e., $d\hat K(Y_0)>0$. On a compact subset of the regular part, the inequality has a positive lower bound. Smooth convergence of $\hat K_{r,c}$ to $\hat K$ preserves it for all sufficiently small $r$ and $|c-\frac92|$. Thus, the only places where we must modify either field are the fixed small Hill-scaled neighborhoods of the saddle-centers.

We now verify the hypotheses of Lemma \ref{lem: two-stage-gluing} at these saddle-centers. The full momentum shear affects the linear coefficients of the primitive difference $\iota_{Y_e-Y_1}\omega_0$ and $\iota_{Y_0-Y_1}\omega_0$. The explicit matrix computation is given in Appendix \ref{app: neck-calculation}.

Near $S_{1,-}$, we choose normal coordinates with $w_3<0$ on the Earth side. Pulling $Y_e$ back through the full shear and rescaling gives
\begin{equation}\label{eq: Earth-primitive}
 \iota_{Y_e-Y_1}\omega_0=dG_e,\qquad
 G_e=\frac1r\big(-A(r,c)w_1-B(r,c)w_4+O(|w|^2)\big).
\end{equation}
At $(r,c)=(0,\frac92)$, its coefficients satisfy $0.520<A<0.522$ and $0.079<B<0.081$.
These bounds imply $A\sqrt{\lambda_h}>|B|\sqrt{\lambda_c}$.
The factor $r^{-1}$ causes no problem because the interpolation lemma (Lemma \ref{lem: two-stage-gluing}) was proved uniformly in $t=r$. The exact coefficients are computed in Appendix \ref{app: neck-calculation}, including the momentum shear. Omitting that shear would change precisely the coefficients on which the strict inequality depends.

On the Moon side,
\begin{equation}\label{eq: radial-primitive}
 \iota_{Y_0-Y_1}\omega_0=dG_0^{\rm loc},\qquad
 G_0^{\rm loc}=d_3(r,c)w_1+d_2(r,c)w_4+O(|w|^2),
\end{equation}
At the limiting saddle, $0.489<d_3<0.491$ and $0.004<d_2<0.005$. The coefficient of the $w_1$ term is therefore much larger than that of the $w_4$ term, and the same domination inequality holds with $t=1$ and $\sigma=1$. Because the inequalities are strict, they remain true for the nearby moving saddle-centers. In this way, the Earth and Moon fields can both be joined to the same local field $Y_1$ by Lemma \ref{lem: two-stage-gluing}.

Let $J_2=\left(\begin{smallmatrix}0&-1\\1&0\end{smallmatrix} \right)$ and $\mathcal Q(y,x)=(J_2y,J_2x).$
The map $\mathcal Q$ is symplectic, $\hat K\circ\mathcal Q=\hat K$, and $\mathcal Q_*Y_0=Y_0$. It sends the two Earth-facing Hill saddle-centers to the two outer ones. Thus, the limiting calculation at the second neck has the same absolute coefficients $d_3,d_2$, and the same strict inequality. Smooth convergence to the Hill limit implies that the strict inequality persists for small $\mu$. The antipodal involution gives antipodal saddle-centers in each neck.

We can now assemble the global Liouville vector field. At the first neck, apply the interpolation lemma on the Earth side and on the Moon side, using the same $Y_1$ in the middle. At the second neck, only the Moon side field has to be joined to the analogous local field as $Y_1$. Away from the neighborhoods, we keep $Y_e$ and $Y_0$ unchanged.

All the modifications are Hamiltonian corrections of Liouville vector fields, so the resulting field $Y$ is Liouville. It is transverse to the entire compact regularized energy region wherever the level is regular. The choices can be made equivariantly under the antipodal involution: the local normal forms and the cutoff functions are transported from each saddle-center to its antipodal one. Consequently, the field $Y$ descends to the regularized quotient. The one-form $\alpha=\iota_Y\omega_0$ is the desired contact form.

\subsection{Dividing spheres and contact structure}
We turn to the dividing spheres. The hypersurface $w_3=0$ cuts the regular energy level in a two-sphere. Its equator is the Lyapunov orbit, and its two open hemispheres are transverse to the flow, in opposite directions. For the foliation criterion, we require that each hemisphere be the projection of a nicely embedded finite energy plane. We first prove this directly. Afterward, we cap the necks with standard contact balls to identify the contact structure on each chamber. This is needed because the foliation criterion (Theorem \ref{thm: criterion of FEF}) is formulated for a punctured standard tight lens space, not just for an arbitrary contact three-manifold with boundary.

We work in the saddle-center normal form coordinates, where
$\hat K=k_i+F(I,J)$, $I=\frac12(w_1^2-w_3^2)$ and
$J=\frac12(w_2^2+w_4^2)$, with $\partial_IF,\partial_JF>0$, as in \eqref{eq: exact-normal-form}.

We keep the original symplectic normalization for the global contact form $\alpha$. If $\Psi_i$ denotes the full map from the rescaled normal form chart at the $i$-th neck to the original coordinates, then $\Psi_i^*\omega_0=\kappa_{r,c}\omega_0$, where $\kappa_{r,c}=\frac83r^2c$. Near the dividing sphere we therefore use the normalized local contact form $\beta:=\kappa_{r,c}^{-1}\Psi_i^*\alpha=\iota_{Y_1}\omega_0$. The local calculations below are made with $\beta$.

\begin{lem}\label{lem: explicit-local-planes}
For every sufficiently small $\delta>0$, the sphere $w_3=0$ in the normal form level $F(I,J)=\delta$ consists of two projections of nicely embedded finite energy planes for a suitable adapted almost complex structure.
\end{lem}

\begin{proof}
Near the dividing sphere $w_3=0$, the Liouville vector field is $Y_1$ from \eqref{eq: local-Y1}, and we write $\beta=\iota_{Y_1}\omega_0$ on this energy level
$F(I,J)=\delta>0$.

Let $v=w_1$ and
$\varrho=(w_2^2+w_4^2)^{1/2}=(2J)^{1/2}$. The energy equation
uniquely determines $J=j(\frac12v^2)$, where $j'=-\frac{\partial_IF}{\partial_JF}<0$. The equator $v=0$ is the Lyapunov orbit, with $j(0)>0$. The poles are $v=\pm v_*$, determined by $j(\frac12v_*^2)=0$.

We first prescribe a compatible complex structure $J_\xi$ on the contact planes $\xi=\ker\beta$ along the two open hemispheres.  Away from the poles, use the angular coordinate $\theta=\arg(w_2+iw_4)$ and the contact frame
\[
 e_1=\partial_v-\frac{\partial_IFv}{\partial_JF\varrho}\partial_\varrho,
 \qquad
 e_2=\partial_{w_3}-\frac{3v}{2J}\partial_\theta.
\]
Here the coordinate vector fields are understood on the energy
surface.  Set $b=\frac32\partial_IFv^2+\partial_JFJ$.  Direct calculation
gives $d\beta(e_1,e_2)=\frac{b}{\partial_JFJ}>0$.  Hence, any smooth function $f(v^2)>0$ defines a compatible almost complex structure on the open hemispheres by
$J_\xi\cdot e_1 =f(v^2)e_2$ and  $J_\xi\cdot e_2=-f(v^2)^{-1}e_1.$
Choose $f=1$ near $v=0$, so that the structures prescribed on
the two hemispheres agree smoothly near the Lyapunov equator. We shall determine $f$ near the poles from the smoothness
condition below.

For either choice of sign for $v$, we seek a plane whose projection covers the corresponding hemisphere, in the form
\[
 u(s,t)=\bigl(v(s),\varrho(s)\cos(2\pi t),0,
\varrho(s)\sin(2\pi t)\bigr),
 \qquad t\in\mathbb R/\mathbb Z,
\]
where $\varrho(s) =\sqrt{2j(\frac12v(s)^2)}$.
We require $v(s)\to\pm v_*$ as $s\to-\infty$ and
$v(s)\to0$ as $s\to+\infty$.  Let $\pi$ denote the projection onto $\xi=\ker\beta$ along the Reeb vector field of $\beta$.  The
projected Cauchy-Riemann equation $\pi \partial_tu=J_\xi\cdot \pi \partial_su$ becomes a scalar equation. Indeed,
$\pi \partial_su=v'(s)\, e_1$ and  $\pi \partial_tu=-\frac{2\pi J\partial_IFv}{b}e_2,$
so that
\begin{equation}\label{eq: hemisphere-ode}
 v'(s)=-\frac{2\pi J\partial_IF}{b\,f(v^2)}v.
\end{equation}
Because the coefficient is negative, solutions on the positive hemisphere decrease towards $0$, while those on the negative hemisphere increase towards $0$. Near $v=0$, the coefficient approaches $2\pi \frac{\partial_IF}{\partial_JF}$ by our choice $f(0)=1$. Thus $v(s)$
approaches the equator exponentially as $s\to+\infty$.

We now choose $f$ near each pole.  In order that the polar
parametrization extends smoothly over the puncture, we require
$\varrho'(s)=2\pi\varrho$ near $\varrho=0$.  Differentiating
$J=j(\frac12v^2)$ and using \eqref{eq: hemisphere-ode}, we obtain
\[
\varrho'(s)=-\frac{\partial_IFv}{\partial_JF\varrho}v'(s)
=\frac{\pi \varrho (\partial_IF)^2v^2}{\partial_JF\,b\,f(v^2)}.
\]
The desired equation holds precisely when
\begin{equation}\label{equ: near the pole}
f(v^2)=\frac{(\partial_IF)^2v^2}{2\partial_JF\,b}>0
\end{equation}
near $v=\pm v_*$. This expression is smooth at the poles, where $J=0$ and  $b=\frac32\partial_IFv_*^2$. Its limiting value is $\frac{\partial_IF}{3\partial_JF}$. We now choose a smooth $f(v^2)>0$ that has this expression near the poles and equals $1$ near the equator.  Equation \eqref{eq: hemisphere-ode} then has the required solution, unique up to translation of $s$.

Let us check the extension at a pole.  In the complex coordinate
$z=e^{2\pi(s+it)}$, the equation $\varrho'(s)=2\pi\varrho$
gives $w_2+iw_4=Cz$ for some $C>0$.  Moreover, the energy
relation determines $v$ as a smooth function of $\varrho^2$
near $v=\pm v_*$.  Therefore $u$ extends smoothly over $z=0$,
and its differential there has rank two.

The compatible structure extends smoothly across each pole. Indeed, the projected hemisphere is transverse to the Reeb field there. Hence $\pi\circ du$ is an isomorphism at the pole. The identity $J_\xi\circ\pi\circ du=\pi\circ du\circ j$ defines a smooth compatible complex structure in a neighborhood of the pole and agrees with the preceding formula away from it. The structures on the two hemispheres also agree near the equator, where $f=1$.

It remains to determine the pseudo-holomorphic planes in the symplectization. Write the lift as $\widetilde u(s,t) =(a(s,t),u(s,t))$. On the dividing sphere, we have $u^*\beta=2\pi J(s)\,dt$. Thus, the remaining two Cauchy-Riemann equations are satisfied by
$\partial_sa=2\pi J(s)$ and $\partial_ta=0$.  Near the poles, $J=\frac12\varrho^2= \frac12C^2|z|^2$, thus $a(s)$ extends smoothly over the puncture. At the other end, $J(s)\to j(0)>0$ and hence
$a(s)\to+\infty$. The plane is positively asymptotic to the
Lyapunov orbit.

Finally, the projected planes are embedded because each parametrizes one open hemisphere. Stokes' theorem gives its Hofer energy with respect to $\beta$ as $2\pi j(0)$. For the original contact form $\alpha$, the corresponding lift is $(\kappa_{r,c}a,\Psi_i\circ u)$, with Hofer energy $2\pi\kappa_{r,c}j(0)=\mathcal A(P_{i,2})$. The compatible structure $J_\xi$ prescribed on the sphere extends first to a neighborhood and then to the remaining contact planes, since the space of compatible almost complex structures is contractible.
\end{proof}

\begin{prop}\label{prop: alpha and J}
For some $\epsilon_0,\mu_*>0$, let $0<\mu<\mu_*$ and $E=L_1+\epsilon\eta^2$, $0<\epsilon< \epsilon_0$, $E\neq L_2$. There exists a contact form $\alpha= \iota_Y\omega_0$ on $\mathcal M^1_{\mu,E}$ induced by a Liouville field $Y$ whose Reeb trajectories are the regularized Hamiltonian trajectories. Each chamber is contactomorphic to a standard tight $\RP^3$ with one or two open balls removed. There are dividing spheres $\mathcal S_1$, and $\mathcal S_2$ when $E>L_2$, and an adapted almost complex structure $J_\xi$ such that each sphere consists of two projections of nicely embedded finite energy planes asymptotic to its Lyapunov equator. The same $J_\xi$ is used on both sides of $\mathcal S_1$, and on one side of $\mathcal S_2$ if $E>L_2$.
\end{prop}

\begin{proof}
We check the required conditions of the foliation criterion in Section \ref{sec: criterion}. First, the Liouville vector field $Y$ constructed above satisfies $d\hat K_{r,c}(Y)>0$ on every regular part of the compact subset $\mathcal M_{\mu,E}^1$. Since $d\alpha=\omega_0$ and $Y$ is transverse, $\alpha= \iota_Y\omega_0$ restricts to a contact form on the energy surface. Its Reeb vector field is a positive multiple of the Hamiltonian vector field. In particular, the periodic trajectories and their stable and unstable manifolds coincide up to reparametrization. The interpolation regions in Lemma~\ref{lem: two-stage-gluing} have fixed size. By the construction in Section \ref{sec: smooth continuation}, choose $\epsilon_0>0$ and then $r_0>0$ sufficiently small. By Lemma~\ref{lem: moving-critical-values}, the normal form excesses $\delta_i=a_i(r,c)(c_i(r)-c)$ are uniformly small whenever a corresponding neck is open. This includes the interval between $L_1$ and $L_2$, since $\eta^{-2}(L_2-L_1)\to0$.

For $\delta_i>0$, the sphere $w_3=0$ in the normal form level has equator $w_1=0$. Its hemispheres are transverse to the flow in opposite directions, because $\dot w_3=\partial_IFw_1$. The equator is the positive hyperbolic Lyapunov orbit with $\mu_{\rm CZ}=2$. The characteristic foliation of the sphere is explicit. In the rescaled normal form chart, the global contact form restricts to $\left.\Psi_i^*\alpha\right|_{w_3=0}=\tfrac{\kappa_{r,c}}2(w_2\,dw_4-w_4\,dw_2)$. There are two elliptic singularities, at the poles, with characteristic leaves running between them along meridians.

It remains to identify the contact structures on each chamber. The argument is a continuation through the critical energy, with the neck removed and replaced by a standard cap. More precisely, cut on the fixed region $|w_3|=a$ for sufficiently small $a>0$ and attach the standard contact balls from the saddle-center model. The two-sided estimates in Section \ref{sec: two-sided estimates} give the same construction on both sides of each critical value. After capping, the singular point is removed, and the capped manifolds and contact forms vary smoothly through the critical value. Below $L_1$ the filled Earth and Moon components are standard tight copies of $\RP^3$ by \cite{AFKP12}. Keep the first neck capped while crossing $L_1$. On the Moon side keep the second neck capped while crossing $L_2$.

At a subcritical energy, the capped components are standard tight $\RP^3$. The capped contact forms vary smoothly through the critical levels, so Gray stability shows that every capped chamber remains standard tight. Removing one or two caps gives a standard tight $\RP^3$ with one or two open balls removed,
as required in the finite energy foliation criterion of Section \ref{sec: criterion}.

Finally, for each open neck, Lemma~\ref{lem: explicit-local-planes} provides the two finite energy planes corresponding to a positive $\delta_i$ under a local compatible almost complex structure $J_\xi$.
The spheres tend to $l_i$ as $E\downarrow L_i$ for fixed $\mu$. Extend $J_\xi$ over the remaining contact planes. The final global complex structure agrees on the two sides of each common dividing sphere and can be chosen invariant under the antipodal involution.
\end{proof}

\section{Retrograde orbits and the proofs of the main results}\label{sec: completion}
We now complete the proofs of the main results. We first identify the retrograde orbits that will serve as binding orbits of the finite energy foliations. We then compare the actions of the two Lyapunov orbits and use the foliations, reversibility, and the local saddle-center dynamics to obtain homoclinic/heteroclinic orbits and periodic orbits.

\subsection{Retrograde orbits}
\label{subsec:retrograde orbits}
The foliation criterion requires a transverse $2$-unknot $P_3$ of rational self-linking number $-\frac12$ in each chamber, together with a $2$-disk in its interior. The disk is needed for the criterion itself, but its area does not play a role here. By the index estimates of Section~\ref{sec: index-estimates}, there are no interior contractible periodic orbits with rotation number $1$. Thus the obstruction set in Theorem~\ref{thm: criterion of FEF} is empty regardless of the area of the disk. We will use the standard argument of \cite[Section~3.1]{LS25}.

%On the Earth side, we take the retrograde family $P^e_{3,\mu,E}$ constructed in Section~\ref{subsec: Rotating Kepler problem}. Fix a target energy in the range under consideration, and continue this orbit from an energy $E_-<L_1(\mu)$ close to the first critical value. This continuation remains in a compact region away from the neck. At $E_-$ the orbit is the usual Birkhoff retrograde orbit, hence a transverse $2$-unknot with rational self-linking number $-\frac12$,  admitting a $2$-disk in the chamber.

On the Earth side, use the smooth retrograde family
$P^e_{3,\mu,E}$ constructed in
Section~\ref{subsec: Rotating Kepler problem}.
For fixed small $\mu>0$, this family continues from an energy
$E_-<L_1(\mu)$ to the energies considered here and stays away
from the necks and the collisions. Using the smooth capped contact family from
Proposition~\ref{prop: alpha and J} and Gray stability, this
continuation gives a transverse isotopy of the subcritical
Birkhoff retrograde orbit. Hence $P^e_{3,\mu,E}$ remains
$2$-unknotted with rational self-linking number $-\frac12$.
A $2$-disk in the capped chamber can be isotoped, relative
to its boundary, to avoid the cap and therefore lies in
the interior of the Earth chamber.

On the Moon side, the existence of such a retrograde orbit must be established uniformly as $\mu\to0$. We do this using the shooting construction for Hill's lunar problem.

\begin{prop}\label{prop: continuation of P_3^m}
For $(r,c)$ in a sufficiently small neighborhood of $(0,\frac92)$, with $r\geq0$, the rescaled problem has a retrograde orbit in the Moon chamber. These orbits can be chosen in a common compact set away from collision and the saddle-centers, with uniformly bounded periods. Each is $2$-unknotted, has rational self-linking number $-\frac12$, and admits a $2$-disk in the interior of the chamber.
\end{prop}
\begin{proof}
The argument relies on a topological crossing rather than on the nondegeneracy of an individual retrograde orbit.

Use the shooting construction of a quarter-retrograde orbit as in \cite[Section~7]{LS26}, which produces two shooting arcs. For the Hill problem with $c$ close to $\frac92$, the two shooting arcs have at least one topological crossing. Proposition~2.11 of \cite{LS26} gives a nearby family of such retrograde orbits. The corresponding trajectories have uniformly bounded time intervals and stay away from collisions and the saddle-centers. These orbits are symmetric with respect to both axes.

The rescaled PCR3BP shooting maps converge in $C^\infty$ to the Hill shooting maps on these compact sets. The Hill problem has an additional symmetry that reduces the shooting construction to a quarter of a retrograde orbit. This symmetry is lost for positive mass ratios, so we use the corresponding double shooting construction, based only on the reversibility that persists in the restricted three-body problem. In Moon-centered coordinates, shoot forward from the negative horizontal semiaxis and backward from the positive horizontal semiaxis, with initial conditions fixed by the remaining reversing involution, to the section over the negative vertical semiaxis. Restrict both shooting maps to compact subarcs containing a Hill topological crossing, with uniformly bounded hitting times. The two maps are perturbed independently; their persistent intersection gives a half-retrograde orbit joining two points of the fixed set, see also \cite[Section~7]{LS26}. We conclude that for small $r>0$, the perturbed arcs still intersect near the topological crossings. This gives a retrograde orbit for every parameter $(r,c)$ in a small neighborhood of $(0,\frac92)$, with one common compact bound and one common period bound. Reflection by $(p_1,p_2,q_1,q_2)\mapsto(-p_1,p_2,q_1,-q_2)$ closes the orbit.

Choose the perturbed shooting intersection in a sufficiently small neighborhood of the selected Hill crossing. The resulting closed orbit is $C^1$-close to the Hill retrograde orbit and has a simple retrograde configuration projection around the Moon. Under the capped contact identifications supplied by Gray stability, sufficiently close transverse embeddings are transversely isotopic. Hence the perturbed orbit is $2$-unknotted and has rational self-linking number $-\frac12$.

Fill the one or two neck balls and take a $2$-disk in the Moon chamber without meeting the neck balls. Since the self-linking number is $-\frac12$, Giroux elimination leaves one positive nicely elliptic singularity. See \cite[Section~3.1]{LS25}. This is the disk required in Theorem~\ref{thm: criterion of FEF}.
\end{proof}

\subsection{The actions of the two Lyapunov orbits}\label{sec: action-order}
We now turn to the two Lyapunov orbits when both necks are open. The symmetry of Hill's problem exchanges these two necks when $r=0$. For positive mass ratios, this symmetry no longer holds. We prove below that the Lyapunov orbit $P_{1,2}$ at the first neck has larger action throughout the two-neck energy range. The actions are independent of the choice of primitive because the periodic orbits are contractible in the local symplectic normal form charts.

\begin{lem}\label{lem: action-order}
After possibly decreasing $\epsilon_0$ and $\mu_*$, for every
$0<\mu<\mu_*$ and every energy
$L_2(\mu)<E<L_1(\mu)+\epsilon_0\eta^2,$ we have $\mathcal A(P_{1,2})>\mathcal A(P_{2,2})$.
\end{lem}

\begin{proof}
Use the unregularized Hill scaling from Section~\ref{sec: Hill's problem} and set
\[
 \bar E=\eta^{-2}\left(E+\frac32-2\mu+\frac{\mu^2}{2}\right),
 \qquad
 \bar L_i(\eta)=\eta^{-2}\left(L_i+\frac32-2\mu+\frac{\mu^2}{2}\right).
\]
Recall that
$\bar L_1(\eta)=-\frac92-\eta+O(\eta^2), \bar L_2(\eta)=-\frac92+\eta+O(\eta^2),$
so that $\bar L_2(\eta)-\bar L_1(\eta)=2\eta+O(\eta^2)$.

Near the $i$-th saddle-center equilibrium, take symplectic normal form coordinates as \eqref{eq: exact-normal-form} in which the scaled Hamiltonian is
$\bar L_i(\eta)+F_i(I,J,\eta)$, where $I$ and $J$ denote the hyperbolic and elliptic actions, respectively. The local Lyapunov family $P_{i,2}$
is given by $I=0$, and its scaled action is
$\bar{\mathcal A}_i(\eta,s)=2\pi J_i(\eta,s)$, where
$
 F_i(0,J_i(\eta,s),\eta)=s-\bar L_i(\eta).
$
Writing
$\omega_{c,i}(\eta):=\partial_JF_i(0,0,\eta)>0,$
we see from the implicit function theorem that $J_i(\eta,s)$, and hence
the period function
$T_i(\eta,s)=\partial_s\bar{\mathcal A}_i(\eta,s)=\frac{2\pi}{\partial_JF_i(0,J_i(\eta,s),\eta)},$
admit smooth extensions to a common neighborhood of
$(\eta,s)=(0,-\frac92)$. In particular,
\[
 \bar{\mathcal A}_i(\eta,\bar L_i(\eta))=0,
% \qquad
% \partial_s\bar{\mathcal A}_i(\eta,s)=T_i(\eta,s),
 \qquad
 T_i(\eta,\bar L_i(\eta))=\frac{2\pi}{\omega_{c,i}(\eta)}>0.
\]
The action in the original symplectic coordinates satisfies
$
 \mathcal A(P_{i,2})=\eta^2\bar{\mathcal A}_i(\eta,\bar E).
$

At $\eta=0$, the symmetry of the Hill problem exchanging the two
saddle-center equilibria identifies the period functions of their local Lyapunov families. Thus $\mathcal{\bar A}_1(0,s)=\mathcal{\bar A}_2(0,s)$ and $T_1(0,s)=T_2(0,s)$ for $s$ near $-\frac92$. By the smooth parameter dependence just established, there are constants $m,C>0$ and a fixed sufficiently small energy neighborhood
such that, for every sufficiently small $\eta>0$, $T_1(\eta,s)\ge m$ and $|T_1(\eta,s)-T_2(\eta,s)|\le C\eta$, whenever the corresponding local Lyapunov orbits exist. The second estimate is needed only for $s\ge\bar L_2(\eta)$.

Using the action-period identity, we obtain
\begin{align*}
 \bar{\mathcal A}_1(\eta,\bar E)
 -\bar{\mathcal A}_2(\eta,\bar E)
 &=\int_{\bar L_1(\eta)}^{\bar L_2(\eta)}T_1(\eta,s)\,ds
   +\int_{\bar L_2(\eta)}^{\bar E}
        \bigl(T_1(\eta,s)-T_2(\eta,s)\bigr)\,ds\\
 &\ge m\bigl(\bar L_2(\eta)-\bar L_1(\eta)\bigr)
       -C\eta\bigl(\bar E-\bar L_2(\eta)\bigr).
\end{align*}
For $\eta$ sufficiently small,
$\bar L_2(\eta)-\bar L_1(\eta)\ge\eta$. Moreover, the energy range gives $0<\bar E-\bar L_2(\eta)<\epsilon_0.$ Choose $\epsilon_0>0$ small enough that the preceding estimates hold
throughout the energy range and $C\epsilon_0<\frac m2$, and then decrease $\mu_*$ if necessary. It follows that
\[
 \bar{\mathcal A}_1(\eta,\bar E)
 -\bar{\mathcal A}_2(\eta,\bar E)
 \ge\eta(m-C\epsilon_0)>\frac{m}{2}\eta>0.
\]
Multiplying by $\eta^2$ proves
$\mathcal A(P_{1,2})>\mathcal A(P_{2,2})$.
\end{proof}

\subsection{Reversibility and connections in the Moon chamber}
\label{sec: reversible-connections}

In the restricted problem, the anti-symplectic involution $\mathcal R$ acts by $\mathcal R(p_1,p_2,q_1,q_2)=(-p_1,p_2,q_1,-q_2)$. It preserves the Hamiltonian and reverses its flow:
\[
  \mathcal R^*\omega_0=-\omega_0,\qquad
  H_\mu\circ\mathcal R=H_\mu,\qquad
  \mathcal R\circ\varphi_t=\varphi_{-t}\circ\mathcal R.
\]
We first explain why the local saddle--center normal forms can be
chosen compatibly with this involution. We then arrange that the
contact form on the Moon chamber is anti-invariant under $\mathcal R$.

Let $l_i$ be either of the two collinear saddle--centers, with critical value $L_i(\mu)$. Since $\mathcal R$ fixes $l_i$, its linearization exchanges the stable and unstable eigendirections and acts as a reflection on the center plane. Therefore, we may choose linear symplectic coordinates in which $\mathcal R$ is represented by
\begin{equation}\label{eq: normal-reverser}
  S(w_1,w_2,w_3,w_4)=(-w_1,w_2,w_3,-w_4).
\end{equation}
In these coordinates, after subtracting $L_i(\mu)$, the Hamiltonian has an expansion
\[
K(w)=K_2(w)+\sum_{k\geq 3}K_k(w),\qquad
K_2(w)=\frac{\lambda_h}{2}(w_1^2-w_3^2)+\frac{\lambda_c}{2}(w_2^2+w_4^2),
\]
where $\lambda_h,\lambda_c>0$ may depend on $i$ and $\mu$, and
$K\circ S=K$. The subsequent analytic terms of order at least $3$ can be obtained without losing this symmetry.

Indeed, at degree $k$ the homological equation is
$\{G_k,K_2\}+\Gamma_k=H_k,$ where $H_k$ is the degree-$k$ term at the current step, $\Gamma_k$
is its resonant part, and $G_k$ has no resonant part. The resonant polynomials are precisely the polynomials in
$I=\frac12(w_1^2-w_3^2), J=\frac12(w_2^2+w_4^2).$
Both $I$ and $J$ are invariant under $S$. Moreover, since $S$ is
anti-symplectic and $K_2\circ S=K_2$, $\{G\circ S,K_2\}=-\{G,K_2\}\circ S.$ Thus pullback by $S$ preserves the resonant and nonresonant subspaces. If $H_k\circ S=H_k$, its resonant part satisfies $\Gamma_k\circ S=\Gamma_k$, while uniqueness of the nonresonant solution gives $G_k\circ S=-G_k$. Because $S$ is anti-symplectic, the Hamiltonian flow of such a $G_k$ commutes with $S$. Consequently, every normalization step is $S$-equivariant. The convergent analytic normalization of \cite[Section~2]{Giorgilli01}
therefore produces local symplectic normal form charts $\Phi_i$ satisfying $\mathcal R\circ\Phi_i=\Phi_i\circ S.$ In particular, the normal form coordinates used below are compatible with the original reversibility.

We next consider the contact form $\alpha$ constructed in
Section~\ref{sec: contact} on the Moon chamber of a regular energy hypersurface $\Sigma_E:=H^{-1}(E)$. By construction, $d\alpha=\omega_0|_{\Sigma_E}$ and $\alpha(X_{H_\mu})>0.$ Near the dividing spheres, this form is anti-invariant: in the local normal form coordinates,
$\Psi_i^*\alpha=\kappa_{r,c}\iota_{Y_1}\omega_0$, where the Liouville field $Y_1$ of
\eqref{eq: local-Y1} satisfies $S_*Y_1=Y_1\circ S$.
Since $S^*\omega_0=-\omega_0$, it follows that
$S^*(\Psi_i^*\alpha)=-\Psi_i^*\alpha$ in these neighborhoods.

To obtain the same property throughout the Moon chamber, define
$\alpha_{\mathrm{rev}} :=\frac12\bigl(\alpha-\mathcal R^*\alpha\bigr).$
Because $\mathcal R$ is an anti-symplectic involution,
\[
  \mathcal R^*\alpha_{\mathrm{rev}}=-\alpha_{\mathrm{rev}},
  \qquad
  d\alpha_{\mathrm{rev}}=\omega_0|_{\Sigma_E}.
\]
Furthermore, reversibility gives
$\mathcal R_*X_{H_\mu}=-X_{H_\mu}\circ\mathcal R$, and hence, at
every point $z$ of the Moon chamber,
\[
\begin{aligned}
  \alpha_{\mathrm{rev},z}(X_{H_\mu}(z))
  &=\frac12\Bigl(
      \alpha_z(X_{H_\mu}(z))
      +\alpha_{\mathcal R z}(X_{H_\mu}(\mathcal R z))
    \Bigr)>0.
\end{aligned}
\]
Thus $\alpha_{\mathrm{rev}}$ is again a contact form defining the
same oriented characteristic trajectories. Since $\alpha$ was
already anti-invariant near the dividing spheres, the averaging
leaves it unchanged there. From now on, we replace $\alpha$ by
$\alpha_{\mathrm{rev}}$ and fix the notation $\alpha$. In
particular,
$
  \mathcal R^*\alpha=-\alpha.
$

Each boundary Lyapunov orbit has one local stable branch and one
local unstable branch lying inside the Moon chamber. Near such an
orbit, choose a small incoming annular section and a small outgoing
annular section, both transverse to the Hamiltonian flow, meeting the
corresponding stable and unstable branches, respectively. The next
lemma gives a sufficient condition for the existence of infinitely
many homoclinic orbits.

\begin{lem}\label{lem: reversible-twist}
Let $P_i,P_j$ be small Lyapunov orbits invariant under $\mathcal R$, possibly equal. Suppose that, on an incoming annulus near $P_j$,
there is an arc of $W^u(P_i)$ admitting a nonconstant real-analytic parametrization up to an endpoint on $W^s(P_j)$. Assume that the remaining points of the arc lie on the side whose local passage returns to the chamber, and that their backward semiorbits remain in the chamber. Then the chamber contains infinitely many geometrically distinct transverse homoclinic orbits to $P_i$.
\end{lem}

\begin{proof}
Choose an incoming annulus $A_j^s$ near $P_j$ and its image
$A_j^u=\mathcal R(A_j^s)$, so close to $P_j$ that the saddle-center
normal form applies throughout the local returning passage. Let $J_j$
be the center action of $P_j$. On both annuli use coordinates
$(\theta,r)$, where $r=\sigma(J-J_j)$, with $\sigma\in\{\pm1\}$ chosen
so that $r>0$ corresponds to the returning side. Thus $r=0$ describes
the local stable circle on $A_j^s$ and the local unstable circle on
$A_j^u$. Choose the angular coordinates compatibly with the reversibility.
The local transition map $l_j:A_j^s\cap\{r>0\}\to A_j^u\cap\{r>0\}$
and the action of $\mathcal R$ between these annuli are then
\begin{equation}\label{eq: reversible-local-map}
 l_j(\theta,r)=(\theta+\Delta_j(r),r),\qquad
 \mathcal R(\theta,r)=(-\theta,r),
\end{equation}
where, after choosing the angular orientation,
\begin{equation}\label{eq: logarithmic-twist}
 \Delta_j(r)=g_j(r)-h_j(r)\log r,\qquad h_j(0)>0.
\end{equation}
Here $g_j,h_j$ are real analytic near $r=0$, see the transition formula in \cite[Section~5]{dPS2}.

Let $\gamma$ be the arc in the hypothesis, parametrized analytically
as $(\theta(s),r(s))$ for $s\in[0,\delta)$, with $r(0)=0$ and $r(s)>0$
for $s>0$. Since $r(s)$ is analytic and not identically zero, it has
the expansion $r(s)=a s^k+O(s^{k+1})$ for some integer $k\geq1$ and
$a>0$. After reducing $\delta$, the function $r(s)$ is strictly
increasing. We may therefore write
$\gamma=\{(f(r),r):0<r<r_0\}$, where
$f(r)=f(0)+O(r^{1/k})$ and $f'(r)=O(r^{1/k-1})$.

By reversibility, $\mathcal R(\gamma)\subset W^s(P_i)\cap A_j^u$,
whereas $l_j(\gamma)\subset W^u(P_i)\cap A_j^u$. In the chosen
coordinates these curves are, respectively, $\theta=-f(r)$ and
$\theta=f(r)+\Delta_j(r)$. Their intersection points satisfy
\begin{equation}\label{eq: homoclinic-intersection-equation}
 2f(r)+\Delta_j(r)=2\pi n,\qquad n\in\mathbb Z.
\end{equation}
Set $G(r)=2f(r)+\Delta_j(r)$. By \eqref{eq: logarithmic-twist},
$G(r)\to+\infty$ as $r\downarrow0$, and
$G'(r)=-\frac{1}{r}h_j(0)+o(\frac{1}{r})<0$ for all sufficiently small $r>0$.
Consequently, for each sufficiently large integer $n$,
\eqref{eq: homoclinic-intersection-equation} has a solution $r_n>0$,
with $r_n\to0$. Since $G'(r_n)\neq0$, the two curves intersect
transversely on $A_j^u$, hence $W^u(P_i)$ and $W^s(P_i)$ intersect
transversely in the energy surface.

Each intersection determines a homoclinic orbit to $P_i$. Its past
follows the backward semiorbit of a point of $\gamma$, its local
passage near $P_j$ lies on the returning side, and its future is the
$\mathcal R$-image of such a backward semiorbit. All these pieces stay
in the chamber. Finally, the annulus $A_j^u$ can be chosen with compact
closure disjoint from $P_i$. Since a homoclinic orbit converges to
$P_i$ in both time directions, it meets this transverse annulus only
finitely many times. The infinitely many intersection points obtained
above therefore belong to infinitely many geometrically distinct
homoclinic orbits.
\end{proof}

\begin{lem}\label{lem: coincident-circle-periodics}
Suppose that the inward unstable circle of a small Lyapunov orbit $P$ is carried onto its whole inward stable circle by a finite-time global transition through the chamber, between local sections transverse to the flow. Then the chamber contains infinitely many geometrically distinct homoclinic orbits to $P$ and infinitely many geometrically distinct periodic orbits.
\end{lem}

\begin{proof}
The coincidence of the two circles gives a one-parameter family of homoclinic trajectories to $P$. These trajectories are geometrically distinct, since each crosses the chosen outgoing section once during its initial departure from a sufficiently small neighborhood of $P$.

Let $C^u\subset A^u$ and $C^s\subset A^s$ denote the unstable and stable circles on small outgoing and incoming annuli, respectively. By the finite-time and transversality assumptions, and the compactness of $C^u$, the global transition extends to a real-analytic local diffeomorphism $G$ on a full annular neighborhood of $C^u$. The restriction $G|_{C^u}$ maps $C^u$ onto $C^s$. Moreover, $G$ preserves the transverse area form $d\alpha$ and is exact: $G^*(\alpha|_{A^s})-\alpha|_{A^u}=d\tau_G$, where $\tau_G$ is its Reeb hitting time. Since $\int_{C^u}\alpha=\int_{C^s}\alpha=\mathcal A(P)>0$, exactness implies that $G|_{C^u}$ has degree $+1$. Being a local diffeomorphism, it is therefore a diffeomorphism, and $G$ restricts to a diffeomorphism between sufficiently small annular collars of the two circles.

Choose compatible angle-action coordinates $(\theta,J)$ on these annuli, with $J=J_P:=\frac1{2\pi}\mathcal A(P)$ on both circles. Set $r=\sigma(J-J_P)$, where $\sigma\in\{+1,-1\}$ is chosen so that $r>0$ is the returning side. In these coordinates the restrictions of $\alpha$ have the form $(J_P+\sigma r)\,d\theta+d\psi$, with possibly different exact terms on the two sections. Since $G$ preserves area and has degree $+1$ on the boundary circle, it preserves the side $r>0$. Thus a lift of $G$ has the form $G(\theta,r) =(g(\theta,r),b(\theta,r))$, where $g(\theta+2\pi,r)=g(\theta,r)+2\pi$ and $b(\theta,r) =a(\theta)r+O(r^2)$, with $a(\theta)>0$.

The local passage $l:A^s\to A^u$ is given by $l(\theta,r)=(\theta+\Delta(r),r)$, where $\Delta(r)=g_0(r)-h(r)\log r$, with $g_0,h$ smooth up to $r=0$ and $h(0)>0$. Hence the return map $F=l\circ G$, defined on a sufficiently small punctured annulus in $A^u$, has a lift $F(\theta,r)=(\Theta(\theta,r),B(\theta,r))$ with $\Theta(\theta,r)=g(\theta,r)+\Delta(b(\theta,r))$ and $B(\theta,r)=b(\theta,r)$. Uniformly in $\theta$, as $r\downarrow0$, we have $\Theta(\theta,r)-\theta\to+\infty$ and $\partial_r\Theta(\theta,r)=-\frac1r h(0)+O(|\log r|+1)<0$.

Fix $r_*>0$ sufficiently small that these properties hold on $S^1\times(0,r_*]$. For every sufficiently large integer $n$ and every $\theta$, the equation $\Theta(\theta,r)=\theta+2\pi n$ has a unique solution $r=r_n(\theta)\in(0,r_*)$. The implicit function theorem shows that $r_n$ is smooth and $2\pi$-periodic, and the logarithmic divergence implies $r_n\to0$ uniformly. Let $\Gamma_n$ be its graph. Along $\Gamma_n$, the angular coordinate is preserved modulo $2\pi$, whereas the radial coordinate becomes $b(\theta,r_n(\theta))$.

We use exactness to show that the radial coordinate is also preserved somewhere on $\Gamma_n$. Since $F$ is a composition of two Reeb transitions, $F^*(\alpha|_{A^u})-\alpha|_{A^u}$ is exact. Integrating over $\Gamma_n$ and using the expression for $\alpha$ above, together with $d\Theta=d\theta$ along $\Gamma_n$, gives
$\int_0^{2\pi}\bigl(b(\theta,r_n(\theta))-r_n(\theta)\bigr)\,d\theta=0.$
Thus $b(\theta_n,r_n(\theta_n))=r_n(\theta_n)$ for some $\theta_n$. The corresponding point is fixed by $F$ and determines a periodic orbit entirely contained in the chamber.

Finally, these periodic orbits are geometrically distinct for infinitely many $n$. Indeed, their intersections with $A^u$ approach $C^u$ as $n\to\infty$, whereas any fixed periodic orbit is compact and disjoint from $C^u\subset W^u(P)$, and hence stays a positive distance from that circle. This completes the proof.
\end{proof}

\begin{prop}\label{prop: reversible-maximal-orbit}
Suppose that a real-analytic reversible Hamiltonian flow on a holed chamber admits a finite energy foliation as in Theorem~\ref{thm: criterion of FEF}. Assume that near each boundary Lyapunov orbit the Hamiltonian has the reversible saddle-center normal form, with $\partial_I F,\partial_J F>0$, and the incoming and outgoing returning annuli of Lemma~\ref{lem: reversible-twist}. Assume also that the reversibility $\mathcal R$ preserves the chamber and has the local expression \eqref{eq: normal-reverser}. If $P_i$ is the unique boundary orbit of maximal action, then the chamber contains infinitely many geometrically distinct homoclinic orbits to $P_i$ and infinitely many geometrically distinct periodic orbits. If the inward stable and unstable branches of $P_i$ do not coincide, then there are infinitely many transverse homoclinic orbits to $P_i$, and the flow has a compact invariant set with positive topological entropy.
\end{prop}

\begin{proof}
Let $W_i^{u,+}$ and $W_i^{s,+}$ denote the unstable and stable branches of $P_i$ directed into the chamber. Choose a regular disk-like page $D_\tau$ of the finite energy foliation close to the inward-pointing boundary hemisphere at $P_i$. For $\tau>0$ sufficiently small, $W_i^{u,+}$ intersects $D_\tau$ in an embedded circle $C_\tau$ bounding a disk $B_\tau\subset D_\tau$. Since $d\alpha$ vanishes on $W^u(P_i)$, the integral of $\alpha$ along every circle transported from $C_\tau$ by the flow equals $\A(P_i)$. In particular, $\int_{B_\tau}d\alpha=\A(P_i)>0$.

We use the page-transport argument of \cite[Corollary~1.4]{LS25}. As $C_\tau$ is carried forward across the families of pages and the rigid cylinders, each transported circle lying in a disk-like page bounds a disk of $d\alpha$-area $\A(P_i)$. If no transported circle met the inward stable circle of a boundary orbit, the structure of the foliation would force infinitely many successive crossings of some rigid cylinder $V$ with negative end at a boundary orbit $P_j$. This is impossible by the finite-area argument, which we recall. A contractible intersection circle in $V$ bounds a disk of area $\A(P_i)$. The disjoint intersection circles arising in the page-transport construction cannot bound nested disks, since such disks would have the same positive area. Hence, infinitely many of them would give infinitely many disjoint positive-area disks in $V$. An essential intersection circle, on the other hand, bounds together with the negative end a half-cylinder of area $\A(P_i)-\A(P_j)$. This area is positive if $j\ne i$, by the unique maximality of $\A(P_i)$. For $j=i$ it would vanish, which is impossible for a half-cylinder in the interior of the transverse rigid leaf. Moreover, two distinct disjoint essential circles would bound nested half-cylinders with the same area, again impossible. Since $V$ has finite $d\alpha$-area, the required infinite sequence of crossings cannot occur. Thus the transported unstable circle meets an inward stable circle of some boundary orbit $P_j$.

Choose such an intersection and a regular disk-like page $D$ sufficiently close to the incoming boundary hemisphere at $P_j$. Denote by $C\subset D$ the transported unstable circle and by $C_j^s=W^s(P_j)\cap D$ the local stable circle. They bound disks $B,B_j\subset D$, respectively, with $\int_Bd\alpha=\A(P_i)$ and $\int_{B_j}d\alpha=\A(P_j)$. Consequently, $\int_Bd\alpha\ge\int_{B_j}d\alpha$, with equality only when $j=i$.

If $C=C_j^s$, equality of the areas forces $j=i$. The whole transported unstable circle then belongs to the inward stable branch of $P_i$, so the two inward branches coincide. The global passage carries the entire inward unstable circle to the inward stable circle, and Lemma~\ref{lem: coincident-circle-periodics} gives infinitely many geometrically distinct homoclinic and periodic orbits in the chamber.

Suppose now that $C\ne C_j^s$. Since $C$ meets $C_j^s$ but $B$ is not strictly contained in $B_j$, there is an arc of $C$ lying outside $B_j$ and approaching a point of $C_j^s$. Follow a sufficiently small part of this arc for a finite positive time to a real-analytic incoming annulus $A_j^s$ in the saddle-center coordinates. This annulus is chosen independently of the page $D$, which need not be real-analytic. The resulting curve is contained in the real-analytic manifold $W^u(P_i)\cap A_j^s$ and has an endpoint on $W^s(P_j)\cap A_j^s$. By $\partial_I F,\partial_J F>0$, the exterior of the stable circle on a sufficiently close incoming page corresponds to $I<0$, or equivalently to the returning side $r>0$ of the local transition. Therefore, we obtain a real-analytic arc $\gamma\subset W^u(P_i)\cap A_j^s$ whose remaining points lie on that side. The local intersection with $A_j^s$ has an analytic parametrization extending through the endpoint. Its transverse coordinate is not identically zero, since $\gamma$ has points with $r>0$, and hence vanishes to finite order at the endpoint. The past trajectories of points of $\gamma$ remain in the chamber, since the arc is obtained by finite-time transport of the inward unstable branch before its first stable intersection.

Lemma~\ref{lem: reversible-twist} now gives infinitely many geometrically distinct transverse homoclinic orbits to $P_i$ contained in the chamber. Since $P_i$ is hyperbolic, the Smale-Birkhoff homoclinic theorem \cite{Smale67} applied to a Poincar\'e return map produces a horseshoe, hence a compact invariant set with positive topological entropy and infinitely many geometrically distinct periodic orbits. This proves both assertions.
\end{proof}

By Lemma~\ref{lem: action-order}, the Lyapunov orbit at the second neck has smaller action than the one at the first. Thus a circle obtained by transporting its local unstable circle may bound a disk strictly inside the stable disk at the first neck, allowing transit to the other chamber.

Choose incoming and outgoing transverse disks $\mathcal D_j^-$ and $\mathcal D_j^+$ near $P_j=P_{j,2}$, bounded by the local stable and unstable circles $C_j^s$ and $C_j^u$. The local transit map $c_j:\operatorname{int}\mathcal D_j^-\to\operatorname{int}\mathcal D_j^+$ is a real-analytic diffeomorphism preserving $d\alpha$. If a disk bounded by the transported circle lies in $\operatorname{int}\mathcal D_j^-$, this map transports the whole disk to the other side.

Globally, we follow the unstable circle through the foliation and analyze its intersections with stable circles; the disk it bounds is defined anew on each planar page.

\begin{prop}\label{prop: outer-lyapunov-homoclinics}
In the setting of Theorem~\ref{thm: FEF for mu small - II}, the compact region $\mathcal M^1_{\mu,E}$ contains infinitely many geometrically distinct homoclinic orbits to $P_{2,2}$. If $\mathcal M^1_{\mu,E}$ contains a heteroclinic orbit between $P_{1,2}$ and $P_{2,2}$, then it contains infinitely many geometrically distinct transverse heteroclinic orbits in each direction between them. The initial heteroclinic orbit need not be transverse.
\end{prop}

\begin{proof}
Set $P_i=P_{i,2}$ and $A_i=\A(P_i)$, so $A_1>A_2>0$ by Lemma~\ref{lem: action-order}. The inward unstable branch of $P_2$ meets a nearby disk-like page in a circle with $\alpha$-integral $A_2$. Follow this circle through the foliation as in \cite[Corollary~1.4]{LS25}. Reeb transitions preserve its $\alpha$-integral. Whenever the transported circle bounds a disk in the current leaf, Stokes gives that disk area $A_2$. The disk is defined on that leaf and need not be the image of the initial disk.

Suppose that no transported circle meets a local stable circle. On an incoming planar page near $P_j$, let $B$ and $B_j^s$ be the disks bounded by the transported and stable circles, with areas $A_2$ and $A_j$. Their disjoint boundaries imply that the disks are disjoint or nested. The inclusion $B_j^s\subset\operatorname{int}B$ is impossible since $A_j\geq A_2$. The inclusion $B\subset\operatorname{int}B_j^s$ requires $j=1$. The local transit map is then defined on all of $B$ and carries its boundary through the first neck. Thus the circle cannot escape through the second neck without a stable intersection.

Indefinite continuation would force infinitely many crossings of one of the finitely many rigid cylinders or dividing hemispheres. The resulting circles are disjoint or identical by uniqueness of the flow. Repetition is impossible: a positive-time return diffeomorphism of a circle would keep its backward iterates away from $P_2$, contrary to their backward asymptotics. Contractible circles bound disks of area $A_2$. Equal areas exclude nesting, so finite area allows only finitely many such circles. An essential circle in a rigid cylinder with negative end $P_j$ would bound a half-cylinder of area $A_2-A_j\leq0$, which is impossible. As in \cite[Corollary~1.4]{LS25}, continuation through a regular family reaches a limiting leaf unless a stable intersection occurs, and transits through the first neck are counted by its dividing hemispheres. Hence a stable intersection must occur.

Let $P_j$ correspond to the first such intersection. If $j=2$ and the transported circle equals the stable circle, it gives a family of homoclinics to $P_2$. Otherwise, for $j=2$, equality of the enclosed areas gives an arc outside the stable disk approaching its boundary. Transporting this arc to an analytic incoming section gives finite-order contact on the returning side. Lemma~\ref{lem: reversible-twist}, applied in $\mathcal M^1_{\mu,E}$, produces infinitely many transverse homoclinics to $P_2$.

We treat $j=1$ after proving the heteroclinic assertion. Take any heteroclinic $\Gamma$ in the hypothesis, oriented from $P_2$ to $P_1$ by reversibility. We first exclude local coincidence along $\Gamma$. Such a coincidence would define a transition $G$ from a maximal connected open subset of the initial local unstable circle of $P_2$ to the final local stable circle of $P_1$, with connecting segments in $\mathcal M^1_{\mu,E}$. This transition is injective, since each orbit has a unique initial point and final approach. If its domain were a proper arc, injectivity would bound the angular variations on both circles. Integrating $G^*\alpha-\alpha=d\tau$, where $\tau$ is the Reeb flight time, would then give finite one-sided limits of $\tau$ at the endpoints. The limiting connecting segments stay in the interior of the region, since they cannot touch a transverse boundary hemisphere or reach a binding orbit in finite time. Transversality of the incoming section extends the transition, and analyticity extends the coincidence, contradicting maximality. Thus the transition is a diffeomorphism between the complete circles. Integrating the same identity gives $A_2=\pm A_1$, contrary to $0<A_2<A_1$.

Transport a small arc of the initial unstable circle of $P_2$ along $\Gamma$ to an analytic incoming annulus near $P_1$. Choose a one-sided subarc $\gamma$ ending at $z\in\Gamma$ on the stable circle. Its contact has finite order, and its backward semiorbits remain in $\mathcal M^1_{\mu,E}$ after shrinking the arc. Write $\mathcal N_-$ for the incoming chamber and $\mathcal N_+$ for the chamber reached by the local passage. Notice that $\mathcal N_-$ and $\mathcal N_+$ coincide in the returning case and distinct in the transit case. The inward branches of $P_1$ in $\mathcal N_-$ cannot coincide, since a point of $\Gamma$ would then belong to both $W^u(P_1)$ and $W^u(P_2)$. Proposition~\ref{prop: reversible-maximal-orbit} gives infinitely many transverse homoclinics to $P_1$ in $\mathcal N_-$.

The local passage sends $\gamma\setminus\{z\}$ to a spiral
$\delta(r)=(\phi(r),r)\subset W^u(P_2)$, where $r=|J-J_1|>0$
on the chosen side. The logarithmic normal form gives
$\phi(r)=-h_0\log r+c+o(1)$ and
$\phi'(r)=-\frac1r h_0+o(\frac1r)$, with $h_0>0$, as in \eqref{eq: logarithmic-twist}. %, see also the proof of \cite[Lemma~4.2]{dPHKS24}.
 If the inward branches of $P_1$ in
$\mathcal N_+$ do not coincide,
Proposition~\ref{prop: reversible-maximal-orbit} gives a transverse homoclinic to $P_1$ there. The successive turns of the spiral approach the
outgoing unstable circle in $C^1$ and hence cross the corresponding
arc of $W^s(P_1)$ transversely infinitely often. The forward semiorbits
of this arc lie in $\mathcal N_+$. This gives heteroclinics from $P_2$ to
$P_1$ and includes the returning case $\mathcal N_+=\mathcal N_-$.

If the passage is a transit $\mathcal N_-\neq \mathcal N_+$ and the inward branches in $\mathcal N_+$ coincide, the global transition obtained as in
Lemma~\ref{lem: coincident-circle-periodics} preserves the transit side and returns the spiral to $\mathcal N_-$ through a second local passage. Put $s=J_1-J>0$ and $t=\phi(s)$ in the logarithmic normal form. Direct local calculation gives $\frac{1}{s(t)}s'(t)\rightarrow -\frac{1}{h_0}$. Expanding the analytic global transition $G(t,s(t))$ as in Lemma \ref{lem: coincident-circle-periodics}, and using the same coefficient $h_0$ in both local passages, the returned curve $(\Theta(t),\rho(t))$ %has the form
%$$\begin{aligned}
%(\Theta(t),\rho(t))&=(g(t)+\Delta (b(t)),b(t))\\
%&=(-h_0\log b(t)+g(t)+g_0(b(t)), a(t)s(t)+O(s^2))
%&=(2t+g(t)-t-h_0\log a(t)+g_0(0)+O(s),a(t)s(t)+O(s^2)).
%\end{aligned}$$}
satisfies $\Theta-F,\Theta'-F'\to0$ and $\rho,\rho'\to0$, where $F(t)=2t+p(t)$ and $p(t)$ is analytic and $2\pi$-periodic. Thus $F$ has only finitely many critical values
modulo $2\pi$. Among the infinitely many distinct departure angles of transverse homoclinics to $P_1$ in $\mathcal N_-$ obtained above, we choose a regular value $\beta$ of $F$. For $t_0$ with $F(t_0)=\beta$, the $C^1$ approximation gives a transverse intersection of the returned curve with the corresponding stable arc of $P_1$ near $t_0+2\pi n$ for every sufficiently large $n$.

All orbit segments lie in $\mathcal M^1_{\mu,E}$. Each heteroclinic visits the chosen outgoing annulus only finitely often, since its closure is compact and disjoint from $P_1\cup P_2$. Thus the intersections give infinitely many geometrically distinct heteroclinics. Reversibility gives the opposite direction. Finally, if the first
stable intersection in the circle construction occurs at $P_1$,
the preceding argument gives transverse connections from $P_2$
to $P_1$. An unstable arc from $P_2$ through this intersection has a returning side whose backward semiorbits remain in $\mathcal M^1_{\mu,E}$. Lemma~\ref{lem: reversible-twist} then gives infinitely many geometrically distinct transverse homoclinics to $P_2$, completing the proof.
\end{proof}

\subsection{Proof of Theorem~\ref{thm: Birkhoff conjecture}}
The subcritical regularized components carry the standard tight contact structure by \cite{AFKP12}. Propositions~\ref{prop: Moon-index} and \ref{prop: Earth-index} show that every contractible periodic orbit on such a component has Conley--Zehnder index at least $3$. The same index estimate holds for periodic orbits in the regular parts of the critical components. Hence the subcritical components are dynamically convex.

A Birkhoff retrograde orbit is transversely isotopic to a Hopf fiber and has rational self-linking number $-\frac12$, as established by the shooting and knot arguments of \cite{Birkhoff1915,LS25}. The rational disk theorem \cite[Corollary~1.8]{HS2} therefore gives the open book in part~(i). The same argument applies to any periodic orbit transversely isotopic to a Hopf fiber.

Let $P'$ be the periodic orbit through an interior fixed point of the first return map to a rational disk page for $P_3$. The page has two disk-like lifts $D_0,D_1$ to $S^3$, each a global surface of section for the lifted flow. If $P'$ is contractible in $\RP^3$, a closed lift would meet the same lifted page at consecutive returns without meeting the other one. This contradicts the global-section property of the other lifted page. Hence $P'$ is noncontractible.

The closed lift $\widetilde P'$ is connected and double-covers $P'$. In one lifted period it meets each of $D_0$ and $D_1$ exactly once. Thus $\widetilde P'$ is a positive one-strand braid about the lifted binding $\widetilde P_3$, and $\widetilde P_3\cup\widetilde P'$ is a positive Hopf link. Consequently, $P'$ is $2$-unknotted with rational self-linking number $-\frac12$, and $P_3\cup P'$ is a Hopf link in $\RP^3$. The annular open book of part~(ii) follows from \cite[Theorem~2.9]{LS26}, which restates the relevant case of \cite[Theorem~1.17]{HSW}.

\subsection{Proof of Theorem~\ref{thm: FEF for mu small - I}}
For $L_1<E<L_2$, only the first neck is open. Proposition~\ref{prop: alpha and J} provides the contact form and the common dividing sphere $\mathcal S_1$. Each chamber is a punctured standard tight $\RP^3$. Its retrograde orbit is $2$-unknotted, has rational self-linking number $-\frac12$ and admits an interior rational disk, as shown in Section~\ref{subsec:retrograde orbits}. Its double iteration is contractible, and Propositions~\ref{prop: Moon-index} and \ref{prop: Earth-index} give $\rho\big((P_3^*)^2\big)>1$.

The Lyapunov orbit $P_{1,2}$ is positive hyperbolic. The saddle-center normal form gives $\rho(P_{1,2})=1$ and $\mu_{\mathrm{CZ}}(P_{1,2})=2$. Its two boundary hemispheres are projections of the local holomorphic planes. Every other interior contractible periodic orbit in either chamber has rotation number greater than $1$. Hence the obstruction set in Theorem~\ref{thm: criterion of FEF} is empty, regardless of the disk-area cutoff. Applying that theorem to both chambers gives two $2-3$ foliations. Since the perturbations of the almost complex structure are supported away from $\mathcal S_1$, the structures agree near the common dividing sphere. Gluing the foliations along the local hemispheres produces the global $3-2-3$ finite energy foliation.

Each chamber has a single boundary Lyapunov orbit. The one-holed multiplicity theorem of \cite{dPS2,LS25} gives infinitely many geometrically distinct periodic and homoclinic orbits to $P_{1,2}$. If the inward stable and unstable branches do not coincide, it also gives infinitely many transverse homoclinics and positive topological entropy.

\subsection{Proof of Theorem~\ref{thm: FEF for mu small - II}}
Let $\tilde\epsilon_0>0$ be a common energy bound for the preceding estimates. For $L_2<E<L_1+\tilde\epsilon_0\eta^2$, Proposition~\ref{prop: alpha and J} provides two dividing spheres $\mathcal S_1$ and $\mathcal S_2$ in $\mathcal M^1_{\mu,E}$. The Earth chamber has one boundary sphere, whereas the Moon chamber has two, and only $\mathcal S_1$ is shared by the chambers. The same proposition provides the contact forms, tight contact structures and local holomorphic planes. The retrograde orbits satisfy the required topological conditions by Section~\ref{subsec:retrograde orbits}. Propositions~\ref{prop: Moon-index} and \ref{prop: Earth-index} exclude interior contractible periodic orbits of rotation number $1$.

Theorem~\ref{thm: criterion of FEF} now gives a $2-3$ foliation in the Earth chamber and a $2-3-2$ foliation in the Moon chamber. Their almost complex structures agree near $\mathcal S_1$, so the foliations glue to a $3-2-3-2$ finite energy foliation of $\mathcal M^1_{\mu,E}$, bounded by $\mathcal S_2$.

In the Earth chamber, the one-holed multiplicity theorem of \cite{dPS2,LS25} gives infinitely many periodic and homoclinic orbits to $P_{1,2}$. In the Moon chamber, Lemma~\ref{lem: action-order} gives $\A(P_{1,2})>\A(P_{2,2})$. Proposition~\ref{prop: reversible-maximal-orbit} therefore gives infinitely many periodic and homoclinic orbits to $P_{1,2}$. If its inward stable and unstable branches do not coincide, there are infinitely many transverse homoclinics and a compact invariant set with positive topological entropy. Finally, Proposition~\ref{prop: outer-lyapunov-homoclinics} gives infinitely many geometrically distinct homoclinic orbits to $P_{2,2}$ in $\mathcal M^1_{\mu,E}$. It also shows that the existence of any heteroclinic orbit between $P_{1,2}$ and $P_{2,2}$ in this region implies the existence of infinitely many geometrically distinct transverse heteroclinic orbits in each direction.

Choose $0<\epsilon_0< \frac{\tilde \epsilon_0}2$ and decrease $\mu_*$ so that $L_2(\mu)-L_1(\mu)<\frac{\tilde\epsilon_0}2\eta^2$. Then
\[
L_2(\mu)+\epsilon_0\eta^2<L_1(\mu)+\tilde\epsilon_0\eta^2,
\]
so all the preceding constructions apply throughout the energy interval stated in the theorem.

\appendix
\section{The coefficients in the neck interpolation}\label{app: neck-calculation}
We verify the coefficients in \eqref{eq: Earth-primitive} and the side and sign conventions used in Lemma~\ref{lem: two-stage-gluing}.

At $S_{1,-}=(-a_{\rm H}^3,0,0,-a_{\rm H})$, the Hessian of \eqref{equ: hat K0}, in the coordinate order $(y_1,y_2,x_1,x_2)$, is
\[
 B=\begin{pmatrix}
 1&0&0&-\sqrt2\\
 0&1&0&0\\
 0&0&3/2&0\\
 -\sqrt2&0&0&-5/2
 \end{pmatrix}.
\]
Set $C_0=2\sqrt7$, $C_1=4+2\sqrt7$ and $C_2=-4+2\sqrt7$, and let $\lambda_h,\lambda_c$ be the positive normal frequencies in \eqref{eq: normal-frequencies}. The following matrix gives symplectic normal coordinates:
\begin{equation}\label{eq: exact-V1}
 V_1=\begin{pmatrix}
 0&\dfrac{-3\sqrt2}{\sqrt{2\lambda_c C_2C_0}}&
 \dfrac{3\sqrt2}{\sqrt{2\lambda_h C_1C_0}}&0\\[4pt]
 \sqrt{\dfrac{\lambda_h C_1}{2C_0}}&0&0&-\sqrt{\dfrac{\lambda_c C_2}{2C_0}}\\[4pt]
 \dfrac{-2\sqrt{\lambda_h}}{\sqrt{C_1C_0}}&0&0&
 \dfrac{-2\sqrt{\lambda_c}}{\sqrt{C_2C_0}}\\[4pt]
 0&\dfrac{\sqrt{C_2}}{\sqrt{2\lambda_c C_0}}&
 \dfrac{\sqrt{C_1}}{\sqrt{2\lambda_h C_0}}&0
 \end{pmatrix}.
\end{equation}
Direct multiplication gives $V_1^TJ_4V_1=J_4$, where $J_4$ is the standard symplectic matrix, and $V_1^TBV_1=\operatorname{diag}(\lambda_h,\lambda_c,-\lambda_h,\lambda_c)$. The linear normal coordinates are obtained by translating the saddle to the origin and applying $V_1$. The subsequent analytic symplectic normalization is tangent to the identity and can be chosen equivariant with respect to the anti-symplectic involution $\mathcal R$. For the perturbed Hamiltonians, one uses smoothly varying symplectic coordinates converging to these normal coordinates. The limiting coefficients, rather than the exact matrix pattern, are what is needed below.

The Earth Liouville primitive is $\lambda_e=-(q_1+\mu)\,dp_1-q_2\,dp_2$. Let $\Psi_{r,c}$ be the elliptic-hyperbolic coordinate change in \eqref{equ:elli-hyper. coordinates}, including the momentum shear and the rescaling \eqref{equ: hat Kr1}. Its conformal symplectic factor, from \eqref{eq: conformal-factor}, is $\kappa_{r,c}=\frac83r^2c$. On a neighborhood of $S_{1,-}$, the following convergence holds with all local derivatives:
\begin{equation}\label{eq: full-Earth-pullback}
\frac{r}{\kappa_{r,c}}\Psi_{r,c}^*\lambda_e\overset{r\downarrow0}{\longrightarrow}-\frac{1}{2\sqrt c}\,d\mathcal Q_e,\qquad
\mathcal Q_e(y,x)=\frac{x_1y_1-x_2y_2}{|x|^2}+\frac12x_1x_2.
\end{equation}
Indeed, expansion of the cotangent matrix gives $\nabla\hat G_0(2\sqrt r\,c^{1/4}x)=\frac12r^{3/2}c^{3/4}\nabla G_0(x)+O(r^{5/2})$, together with $p_1=\frac43r\sqrt c\,\mathcal Q_e+O(r^2)$ and $q_1+\mu=1+O(r)$. The remaining term $q_2\,dp_2$ is of higher order by the same rescaling. Substitution and the value of $\kappa_{r,c}$ give \eqref{eq: full-Earth-pullback} in the local smooth topology.

At the saddle, $\nabla\mathcal Q_e=(0,\frac1{a_{\rm H}},-\frac{3a_{\rm H}}2,0)$ and $\nabla\mathcal Q_e\,V_1=(d_1,0,0,d_4)$, where $d_1,d_4>0$. Thus the limiting coefficients in \eqref{eq: Earth-primitive} are
\begin{equation}\label{eq: exact-A-B}
\begin{split}
 A(c)&=\frac{\sqrt{\lambda_h}}{2\sqrt c\sqrt{2C_1C_0}}
       \left(\frac{C_1}{a_{\rm H}}+3\sqrt2a_{\rm H}\right),\\
 B(c)&=\frac{\sqrt{\lambda_c}}{2\sqrt c\sqrt{2C_2C_0}}
       \left(-\frac{C_2}{a_{\rm H}}+3\sqrt2a_{\rm H}\right).
\end{split}
\end{equation}
At $c=\frac92$, direct evaluation gives $A\simeq0.520598$ and $B\simeq0.079625$. The exact formulas give $0.520<A<0.522$, $|B|<0.081$, $1.773<\lambda_h<1.775$ and $1.464<\lambda_c<1.466$. In particular, $A\sqrt{\lambda_h}>|B|\sqrt{\lambda_c}$. This inequality persists for nearby parameter values. The shear contribution $\tfrac12x_1x_2$ in \eqref{eq: full-Earth-pullback} is essential for these coefficients.

For the Liouville field $Y_0$, the constant term in centered linear normal coordinates is determined by $\frac12V_1^{-1}S_{1,-}=(0,d_2,-d_3,0)$, where $d_2\simeq0.004701$ and $d_3\simeq0.489861$. The exact matrix formulas give $0.004<d_2<0.005$ and $0.489<d_3<0.491$. These bounds imply condition \eqref{eq: domination}. Before the nonlinear normalization, the local expression of the corresponding quadratic primitive is $G_0^{\mathrm{lin}}=d_3w_1+d_2w_4-w_1w_3$. Since the analytic normalizing transformation is tangent to the identity, its effect on this function begins at quadratic order. In the final normal coordinates, $G_0^{\mathrm{loc}}=d_3w_1+d_2w_4+O(|w|^2)$.

At the other Hill saddles, use the symplectic quarter-turn $\mathcal Q(y,x)=(Ry,Rx)$, where $R=\left(\begin{smallmatrix}0&-1\\1&0\end{smallmatrix}\right)$. The limiting Hamiltonian and $Y_0$ are invariant under $\mathcal Q$, and $\mathcal Q(S_{1,-})=S_{2,+}$ and $\mathcal Q(S_{1,+})=S_{2,-}$. Conjugating the normal coordinates by $\mathcal Q$ therefore gives the same absolute values of the relevant linear coefficients and the same domination inequality at those saddles. For sufficiently small $r>0$, the conclusion follows by smooth convergence to the limiting model, without assuming an exact quarter-turn symmetry of the perturbed Hamiltonian. The antipodal involution treats the corresponding antipodal points.

\hspace{3mm}

\noindent\textbf{Acknowledgements}:
LL is partially supported by the National Natural Science Foundation of China (Grant No.~12401238) and the Natural Science Foundation of Shandong Province (Grant No.~ZR2024QA188). LL thanks the School of Mathematics at Shandong University for its support. PS is partially supported by the National Natural Science Foundation of China (Grant No.~W2431007). LL and PS thank the Shenzhen International Center for Mathematics--SUSTech for its support.

\end{document}